\documentclass{JHEP3}
\usepackage{amsmath}
\usepackage{epsfig}

\newcommand{\Ztwo}{{\mathbb{Z}_2}}

\newcommand{\One} {{\bf 1}} 

\newcommand{\tQ}{{\tilde Q}}

\newcommand{\cY} {{\cal Y}}

\newcommand{\cC} {{\cal C}}
\newcommand{\cS} {{\cal S}}

\newcommand{\cz} {{\mathbb Z}}

\newcommand{\cO} {{\cal O}}
\newcommand{\al} {{\alpha}}
\newcommand{\la} {{\lambda}}
\newcommand{\La} {{\Lambda}}
\newcommand{\ep} {{\epsilon}}

\newcommand{\ket}[1] {\left|#1\right>}

\title{Superstring field theory in the democratic picture}

\author{Michael Kroyter\\
Center for Theoretical Physics\\
Massachusetts Institute of Technology\\
Cambridge, MA 02139, USA\\
\\ and\\ \\
School of Physics and Astronomy\\
The Raymond and Beverly Sackler Faculty of Exact Sciences\\
Tel Aviv University, Ramat Aviv, 69978, Israel\\ \\
\email{mikroyt@mit.edu}, \email{mikroyt@tau.ac.il}
}

\abstract{We present a new open superstring field theory, whose string
fields carry an arbitrary picture number and reside in the large Hilbert
space. The redundancy related to picture number is resolved by treating
picture changing as a gauge transformation. A mid-point insertion is
imperative for this formalism. We find that this mid-point insertion must
include all multi-picture changing operators. It is also proven that this
insertion as well as all the multi-picture changing operators are zero
weight conformal primaries.

This new theory solves the problems with the Ramond sector
shared by other RNS string field theories, while naturally
unifying the NS and Ramond string fields. When partially gauge fixed, it
reduces in the NS sector to the modified cubic superstring field theory.
Hence, it shares all the good properties of this theory, e.g., it has
analytical vacuum and marginal deformation solutions.

Treating the redundant gauge symmetry using the BV formalism
is straightforward and results in a cubic action with a single string
field, whose quantum numbers are unconstrained.
The generalization to an arbitrary brane system is simple and includes
the standard Chan-Paton factors and the most general string field consistent
with the brane system.
}

\preprint{MIT-CTP-4037\\TAUP-2900-09}

\begin{document}

\section{Introduction}

The existence of picture number is a central characteristic of the RNS
formalism of superstring theory~\cite{Friedan:1985ge}. Picture number
complicates the construction of RNS string field theories\footnote{Various
steps towards the
construction of superstring field theory can be found
in~\cite{Witten:1986cc,Witten:1986qs,Wendt:1987zh,Preitschopf:1989fc,
Arefeva:1989cm,Arefeva:1989cp,Berkovits:1995ab,Berkovits:2000hf,
Berkovits:2001im,Arefeva:2002mb,Michishita:2004by,Berkovits:2005bt,
Berkovits:2009gi,Kroyter:2009zj,Kroyter:2009zi,Kroyter:2009bg}.
A review of recent progress in string field theory, whose section 8 includes
an introduction to superstring field theory is~\cite{Fuchs:2008cc}.}.
The fact that the pictures of the NS fields are integer, while those of the
Ramond fields are half-integer implies that one should either put a
relative picture changing operator between the two sectors, as is done
in the modified theory~\cite{Preitschopf:1989fc}, or use two (or more)
Ramond fields with a constraint, as in the non-polynomial version of the
theory~\cite{Michishita:2004by}.
Both these resolutions are problematic. The former suffers from collisions of
picture changing
operators~\cite{Wendt:1987zh,Kroyter:2009zi}\footnote{Witten's theory also
does not seem to support some desired classical
solutions~\cite{DeSmet:2000je}.}, while the
constraint of the latter cannot be derived from a covariant action.
We review the current status of RNS superstring field
theories, focusing on the cubic theories, in section~\ref{sec:SSFTintro}.

In this work we propose a new open RNS string field theory,
which does not suffer from these problems.
Our starting point in defining this action relies on two observations made
in~\cite{Berkovits:2001us}. The first is that the cohomology of $Q$ at any
given picture and ghost numbers in the small Hilbert space is the same as the
cohomology of $Q-\eta_0$ at the same picture and ghost numbers in the large
Hilbert space. The
second observation is that relaxing the constraint of a given picture number
does not change the cohomology, unlike the case of $Q$ in the small Hilbert
space, where considering two picture numbers at once results in a
doubling of the cohomology. We explain these issues in
section~\ref{sec:QetaPXi}.

The theory itself is introduced in section~\ref{sec:SSFT}.
The action is cubic and is defined in the large Hilbert space,
\begin{equation}
\label{action}
S=-\oint \cO \Big(\frac{1}{2}\Psi \tQ\Psi+\frac{1}{3}\Psi^3\Big).
\end{equation}
Here, $\Psi$ is the string field and we abbreviated
\begin{equation}
\label{Qtilde}
\tQ\equiv Q-\eta_0\,.
\end{equation}
The string fields are multiplied, as always,
using Witten's star product~\cite{Witten:1986cc}, which we keep implicit.
The integration symbol appearing in~(\ref{action}) is used throughout this
paper to represent the
CFT expectation value in the large Hilbert space, while the standard
integration symbol represents the expectation value in the small Hilbert
space. The mid-point insertion $\cO$ is defined
using a superposition of all multi-picture-changing operators.
We introduce these operators and discuss their properties. In particular, it
is proven that they can always be represented by primary fields.

Another novel property of the action~(\ref{action}), which we present in
section~\ref{sec:Ramond}, is that the NS and Ramond sector
fields are simply unified into a single string field,
\begin{equation}
\Psi\rightarrow \Psi+\al\,.
\end{equation}
Now, $\Psi$ is the NS string field, which is formed by a linear combination
of string fields with arbitrary integer picture number and $\al$ is the
Ramond string field, which is formed by a linear combination
of half-integer picture numbers. Expanding the action~(\ref{action}) in
terms of the new $\Psi$ and $\al$ gives,
\begin{equation}
S= -\oint \cO\Big(\frac{1}{2}\Psi \tQ \Psi+\frac{1}{3}\Psi^3
+\frac{1}{2}\al \tQ \al+\Psi \al^2\Big).
\end{equation}
This is very similar in form to the usual cubic RNS superstring field theory
action~\cite{Preitschopf:1989fc}. The most striking difference being
the fact that the same insertion $\cO$ multiplies all terms in the action.
The consistent inclusion of the Ramond sector enables the covariant
description of all sectors of general D-brane systems in the new formalism.

The field--antifield (BV) formulation of the theory is presented in
section~\ref{sec:BV}, where we also comment on gauge fixing.
The supersymmetry properties of the theory are discussed in
section~\ref{sec:SUSY}.
Conclusions and some open problems are presented in section~\ref{sec:conc}.

Note added: While this work was nearing completion, I learned of the
work~\cite{SchnablGrassi}, which has some similarities to our construction.
It would be interesting to understand the interrelation, if any, between
the two theories.

\newpage
\section{Cubic superstring field theory}
\label{sec:SSFTintro}

In this section we recall the construction of the existing
cubic RNS string field theories and their problems.

The first proposal for such a theory was presented by
Witten~\cite{Witten:1986qs}, following his bosonic
theory~\cite{Witten:1986cc}. While the structure in the two cases is almost
the same, there was one essential new feature that had to be addressed,
namely the picture number. Since in the small Hilbert space, where the
theory was constructed, CFT expectation values are non-zero for picture
number $-2$, it seemed quite sensible to define the NS string field to carry
the ``natural'' $-1$ picture, in order to get a standard form for the kinetic
term in the action.

The interaction term, on the other hand, had to be appended with a +1
picture number, in order to allow for a non-trivial result. This was
achieved by an explicit insertion of the picture-changing operator
$X$ in the action. The only consistent way of inserting the picture changing
operator was as a mid-point insertion. Any other choice would have destroyed
the associativity of the star product and the gauge invariance of the action.
However, it was soon realized that the facts that the mid-point is invariant
under the star product and that the OPE of $X$ with itself is singular,
imply the emergence of singularities that ruin the consistency of
the theory~\cite{Wendt:1987zh}.

The resolution of~\cite{Preitschopf:1989fc,Arefeva:1989cp} (henceforth, ``the
modified theory'') was to change the
picture number of the NS string field to zero and insert an overall factor
of $Y_{-2}$ in front of the action. The action of the NS sector of this
construction is given by,
\begin{equation}
\label{NSaction}
S_{NS}=-\int Y_{-2}\big(\frac{1}{2}\Psi Q\Psi+\frac{1}{3}\Psi^3\big)\,.
\end{equation}
Now, the gauge transformation does not
contain any picture changing operators and in perturbation theory
the factors of $Y_{-2}$ come (at least for trees) next to factors of
$\big(Y_{-2}\big)^{-1}$ from the propagator. Therefore, singularities do not
emerge.

While this solved the problem with picture changing collisions, other
criticism on this theory remained. A common claim was that the equation
of motion derived from the action is not the one needed, due to the
non-trivial kernel of $Y_{-2}$,
\begin{equation}
Y_{-2}(Q\Psi+\Psi^2)=0 \quad \stackrel{?}{\Longleftrightarrow}
 \quad Q\Psi+\Psi^2=0\,.
\end{equation}
In fact, not only the kernel of $Y_{-2}$, but also the space of operators
whose OPE with $Y_{-2}$ is singular, is potentially problematic. However,
these problems may arise only for string fields having these types of
operators as local insertions at the string mid-point. String fields of this
sort suffer from problems regardless of the existence of the mid-point
insertion in the action and (at least) most of them should
be discarded from the construction of the space of string
fields\footnote{See, for example, section 2 of~\cite{Kroyter:2009zj} for
further related discussion. Also note that a
cubic theory, which avoids insertions of operators with a
non-trivial kernel, exists and it seems that it is classically
equivalent to the modified theory~\cite{Berkovits:2009gi,Kroyter:2009zj}.
However, this formalism does not solve the problems with the Ramond sector.
Moreover, the space of operators whose OPE with the insertion of this theory
is singular, is non-trivial. Thus, even if one insists, in contrast to what
we suggest here, that mid-point insertions should be allowed, this
theory is still in no way better than the modified one.}.
While the construction of a space of string fields is generally lacking,
it does not seem that it would be more complicated in the case of the
modified theory. We therefore believe that this
issue does not form a ground for discarding the theory.
Another issue of this theory is the abundance of possible $Y_{-2}$
insertions. However, it was shown in~\cite{Kroyter:2009bg} that
theories based on different choices of $Y_{-2}$'s are classically 
equivalent. Hence, this is also not much of an obstruction to the theory.
Moreover, cubic superstring field theory is very successful in many ways.
In particular, analytical solutions describing the tachyon vacuum and
marginal deformations are known in the
theory~\cite{Erler:2007xt,Aref'eva:2008ad,Fuchs:2008zx}.

There is, however, one serious problem with this formalism related to the
inclusion of the Ramond sector. The Ramond sector carries half-integer
pictures. Hence, it is impossible (with a Ramond field of a well defined
picture) to write an interacting action with a common $Y_{-2}$ factor.
The proposal of~\cite{Preitschopf:1989fc} was to generalize the
action~(\ref{NSaction}) to,
\begin{equation}
\label{PTYaction}
S=-\int \Big(Y_{-2}\big(\frac{1}{2}\Psi Q\Psi+\frac{1}{3}\Psi^3\big)+
  Y\big(\frac{1}{2}\al Q \al+\Psi \al^2\big)\Big),
\end{equation}
where $\al$ is the Ramond field.

That this formulation is problematic can be observed already when deriving
the equations of motion from the action~(\ref{PTYaction}).
These take the form (after acting on them with picture changing operators),
\begin{subequations}
\label{CubEOM}
\begin{eqnarray}
\label{eomCubA}
Q \Psi + \Psi^2 + X \al^2 &=& 0\,,\\
\label{eomCubAl}
Q \al + [\Psi,\al] &=& 0\,.
\end{eqnarray}
\end{subequations}
The first of these equations implies that mid-point insertions are to be
allowed for at least one of the string fields $\Psi$ and $\al$, since
otherwise the Ramond sector becomes trivial.
It is not clear what sort of an operator could have been inserted on
$\al$, in order to produce a $Y$ (that is needed to cancel the $X$) on
$\al^2$. Hence, we assume
that the insertion is on $\Psi$. We cannot assign $X$ to $\Psi$, since then
the term $\Psi^2$ would be singular. Hence, we decompose $\Psi$ as,
\begin{equation}
\Psi=\Psi_0+\xi \Psi_1\,,
\end{equation}
with $\Psi_0$ being an odd component and $\Psi_1$ being an even
one~\cite{Kroyter:2009zi}.
This still does not solve the problem, since singular terms of the form
$\xi X$ from the $\Psi Q\Psi$ term appear now in the action.
One may claim that the $X$ should be canceled against ``half the $Y_{-2}$''
before it is multiplied by the $\xi$.
This means, however, that all string fields in the theory should be
regularized somehow, already at the classical level. In fact, the theory
itself should be regularized.

A similar observation can be made by considering the gauge transformation
of this theory,
\begin{subequations}
\label{cubGauge}
\begin{eqnarray}
\delta \Psi &=&   Q\La+[\Psi,\La]+X[\al,\chi]\,,\\
\delta \al &=& Q\chi+[\al,\La]+[\Psi,\chi]\,,
\end{eqnarray}
\end{subequations}
where $\La$ is the NS gauge string field and $\chi$ is the Ramond
gauge string field, both of which are even string fields.
While the second equation is benign, the first one explicitly shows
that an $X$ mid-point insertion should be allowed for $\Psi$,
as it is not clear how one could avoid it without moving the problem to
$\al$. In fact, when iterated,~(\ref{cubGauge}) inevitably leads to
singularities due to collisions of $X$~\cite{Kroyter:2009zi}.

It might be possible to regularize the theory
by moving the local insertions away from the mid-point. However,
such a regularization would break the associativity of the star product and
the gauge invariance of the theory.
It is not clear whether those could be restored in the limit in which the
regularization is removed, nor is it clear whether the singularities can be
avoided in this limit.
Another option would be to look for an insertion-free
formulation. Such a formulation exists, namely
the non-polynomial theory~\cite{Berkovits:1995ab,Michishita:2004by}.
Here, however, one faces another problem. The theory is defined using two
constrained Ramond fields. If one insists on a covariant formulation of the
theory, the constraint cannot be derived from an action.

One might try to use two Ramond fields in the description of the cubic
theory, in analogy with the situation in the non-polynomial theory. While the
resulting theory has some nice properties, a constraint relating the two
Ramond fields should be introduced~\cite{Kroyter:2009zi}. Nonetheless,
it is not understood how to derive a proper constraint from an action
without introducing explicit mid-point insertions. Thus, the problems
described above persist.
We conclude that a more fundamental revision of the theory is needed.
We derive a cubic theory with the desired properties in
section~\ref{sec:SSFT}.

\section{Cohomology in the large Hilbert space}
\label{sec:QetaPXi}

In this section we review the various representations of vertex operators
in the RNS formalism.
Vertex operators are important to us, since string fields are their
off-shell generalizations. Different choices of
representations of the vertex operators naturally lead to different
string field theory formulations. While in their most familiar
representation, the RNS vertex operators live in the small Hilbert space,
some other useful representations use the large Hilbert space.

The large Hilbert space consists of two copies of the small Hilbert space,
with one of the copies being multiplied by $\xi_0$ \cite{Friedan:1985ge}.
Within the copy without the $\xi_0$ insertion, i.e., within the small
Hilbert space, the on-shell condition for a vertex operator $V$ gets the
form,
\begin{equation}
\label{smallOS}
Q V=0\,.
\end{equation}
The requirement of being in the small Hilbert space is enforced by,
\begin{equation}
\label{smallOsEta}
\eta_0 V=0\,.
\end{equation}
To these relations one has to add the identification of states that differ by
an exact element,
\begin{equation}
\label{cohoEquiv}
V\approx V+Q\La\,,
\end{equation}
where $\La$ also obeys~(\ref{smallOsEta}).
This is the way that the standard cohomology problem is formulated in the
large Hilbert space.
The cubic theories described in section~\ref{sec:SSFTintro}
are off-shell extensions of this representation of the
vertex operators.

If one wishes to consider the second copy of the small Hilbert space, i.e.,
the one with the $\xi_0$ insertion, the on-shell condition
for a vertex operator $V$ can be written as,
\begin{equation}
\label{largeOS}
Q\eta_0 V=0\,.
\end{equation}
Here $\eta_0$ removes the $\xi_0$ insertion and the remaining equation is
the same as~(\ref{smallOS}).
Two solutions should be considered equivalent if they differ by a term of the
form,
\begin{equation}
\label{XicopyEquiv}
\delta V=\xi_0 Q\eta_0 \La\,.
\end{equation}
This relation mimics~(\ref{cohoEquiv}) for the $\xi_0$ copy of the small
Hilbert space. Hence,~(\ref{largeOS}) and~(\ref{XicopyEquiv}) correctly
define the equivalence classes in this space, despite the fact that they
do not look like a cohomology problem.
The expression~(\ref{largeOS}) can also be used for the
whole large Hilbert space. To that end we have to append to the equivalence
relation~(\ref{XicopyEquiv}) the equivalence to zero of the small
Hilbert space. Changing the basis of equivalence generators this can be
written as,
\begin{equation}
\label{2LinGaugeTrans}
\delta V=Q\La_Q+\eta_0 \La_\eta\,.
\end{equation}

There is a natural correspondence between $\xi_0$-based states at ghost and
picture numbers $g$ and $p$ and small Hilbert space states at ghost number
$g+1$ and picture number $p-1$. The change of quantum numbers comes from
using $\eta_0$ and $\xi_0$ as the canonical isomorphism mappings.
Hence, NS string fields in the ``natural'', $p=-1$, $g=1$ picture,
correspond to $p=g=0$ string fields that extend the vertex operators
represented by~(\ref{largeOS}) and~(\ref{2LinGaugeTrans}). While the former
case leads to Witten's theory~\cite{Witten:1986qs},
which suffers from divergences, the later naturally leads to the
non-polynomial theory of Berkovits~\cite{Berkovits:1995ab}, in which these
problems do not arise.

One might wish at this stage to consider also the cohomology of $Q$
in the large Hilbert space. Another possibility would be to consider the
cohomology of $\eta_0$, since using\footnote{Here and elsewhere,
$[A,B]$ stands for the graded commutator, e.g.,
$[Q,\eta_0]\equiv Q\eta_0+\eta_0 Q$.}
\begin{equation}
\label{Qeta0}
[Q,\eta_0]=0\,,
\end{equation}
one realizes that~(\ref{largeOS}) and~(\ref{2LinGaugeTrans}) are exactly
symmetric upon interchanging $Q$ and $\eta_0$. However, in the large Hilbert
space the cohomology of both operators is trivial, as we turn now to show
while further illustrating the symmetry between both operators\footnote{The
symmetry between $Q$ and $\eta_0$ was noted already
by Berkovits and Vafa~\cite{Berkovits:1994vy}. In the $N=4$ language
$J_B$ and $\eta$ are the currents $G^+$ and $\tilde G^+$ respectively.
Also, note that one can completely exchange the roles
of $Q$ and $\eta_0$ and define a ``dual small Hilbert space'', by
considering the $Q$-closed subspace of the large Hilbert space
and studying the cohomology of $\eta_0$ in this space.
This cohomology is the same as that of $Q$ in the ``ordinary small Hilbert
space'', since both ($Q$'s cohomology in one space and $\eta_0$'s
in the other) are relative cohomologies that are defined in the same way in
the large Hilbert space: $\eta_0 V=Q V=0$, $V\approx V+Q\eta_0\La$.}. 
For an arbitrary derivation $d$, the triviality of its cohomology can most
easily be shown if a state $A$ exists such that $dA=\One$, where $\One$ is
the identity element of the algebra on which $d$ acts\footnote{Such a state
is called a ``contracting homotopy''. In the
context of string field theory this structure was used for proving that
the cohomology around Schnabl's solution~\cite{Schnabl:2005gv} is
trivial~\cite{Ellwood:2006ba} (see
also~\cite{Ellwood:2001ig,Erler:2007xt,Fuchs:2008zx}).}. The proof is then
straightforward. Let $V$ be a closed state, i.e., $dV=0$, then $V$ is exact.
Specifically, $V=d(AV)$, since
\begin{equation}
d(AV)=(dA)V+(-)^{(d)(A)}AdV=\One V+0=V\,,
\end{equation}
where $(d)$ and $(A)$ in the exponent stand for the parities of the
derivation and the state $A$.
Such a state exists for $\eta_0$. In fact, a family of
such states exists, namely $\xi(z)$. It is less trivial to find it,
but a similar family exists for $Q$~\cite{NarganesQuijano:1988gb},
namely\footnote{It is easy to read from $P$ the part of $Q$, which has
$\phi$-momentum two. However, the complete $Q$ is related to this part by
a similarity transformation~\cite{Acosta:1999hi} and the generator of this
transformation leaves $P$ invariant.}
\begin{equation}
P(z)=-c\xi\partial\xi e^{-2\phi}(z)\,.
\end{equation}
Hence, we write,
\begin{subequations}
\label{QP1EtaXi1}
\begin{align}
\label{QP1}
Q P(z) &=1\,,\\
\label{EtaXi1}
\eta_0 \xi(z) &=1\,,
\end{align}
\end{subequations}
where $1$ here stands for $\One(z)$, the insertion of unity at $z$,
which does not change the state on which it acts and is of course
$z$-independent.
This completes the proof of the triviality of $Q$ and $\eta_0$.

It is also interesting to consider,
\begin{subequations}
\begin{align}
\label{X}
X(z)&\equiv Q \xi(z)= \big(c\partial\xi
  +e^\phi G_m+e^{2\phi}b\partial \eta
   +\partial(e^{2\phi}b \eta)\big)(z)\,,\\
Y(z)&\equiv \eta_0 P(z)= \big(c\partial\xi e^{-2\phi}\big)(z)\,.
\end{align}
\end{subequations}
Here, $G_m$ is the superconformal matter generator, which in flat background
takes the form\footnote{We use the same conventions as
in~\cite{Fuchs:2008cc}.},
\begin{equation}
G_m=i\psi_\mu \partial X^\mu\,.
\end{equation}
The expression~(\ref{X}) is, however, universal and holds regardless
of the existence of a specific background.

These definitions imply that all the quantum numbers of $X$ and $Y$ are
trivial except for their picture numbers, which are $1$ and $-1$
respectively. In particular, they are zero weight conformal primaries.
These operators are also $Q$-closed. For $X$ it follows from
the fact that it is exact (in the large Hilbert space),
while for $Y$ it follows from the (graded) Jacobi identity and the
relations~(\ref{Qeta0}) and~(\ref{QP1}).
Similarly, these operators are closed with respect to $\eta_0$. All in all
we can write,
\begin{equation}
\label{XYQeta}
QX=QY=\eta_0 X=\eta_0 Y=0\,.
\end{equation}
It is also possible to write $Y$ explicitly as an exact
state in the large Hilbert space,
\begin{equation}
\label{YcY}
Y=Q\cY\,.
\end{equation}
This statement is in a sense trivial, since we already proved that it is
closed and any closed
state can be written in the large Hilbert space as an exact state using $P$.
However, $P$ has a singular OPE with $Y$, which complicates the explicit
construction.
The most straightforward resolution of the OPE singularities would have been
the replacement of the operator product by a normal ordered product. This
strategy does not work, since it leads to a vanishing result, due to
the zeros at the $bc$ and $\xi\eta$ sectors.
We can remedy that by defining $\cY$ as the leading regular term
in the OPE,
\begin{equation}
\cY_0(w)\equiv \oint \frac{dz}{2\pi i} \frac{P(z)Y(w)}{z-w}\,.
\end{equation}
This is almost what we want. This operator is local and
obeys~(\ref{YcY}). However, it has some non-trivial quantum numbers
other than the needed ghost and picture numbers, meaning that, while it
carries zero conformal weight, it is not a primary conformal field.
Since this property will turn out to be of importance to us, we would like
to suggest a conformal primary candidate for $\cY$,
\begin{equation}
\label{YetaPot}
\cY=\frac{1}{5}\,c\,\xi\partial\xi e^{-3\phi}G_m
  -\xi e^{-2\phi}\,.
\end{equation}
Similarly, $X$ can be written, in the large Hilbert space, as a local
$\eta_0$ exact state.

From the discussion above it follows that,
\begin{equation}
\label{PYxiX}
P=\xi Y\,,\qquad \xi=PX\,.
\end{equation}
The last two equalities in~(\ref{XYQeta}) imply that $X$ and $Y$ reside in
the small Hilbert space.
We can now conclude that these operators are ``picture changing operators'',
namely that they define homomorphisms between the cohomologies (in the small
Hilbert space) of $Q$ at picture numbers $p$ and $p\pm 1$.
They are also each other's inverse in the sense of the OPE,
\begin{equation}
X(z)Y(0)\sim 1\,,
\end{equation}
as is implied by~(\ref{PYxiX}).
Nevertheless, as already stated, these operators suffer from singularities
in their OPE's with themselves,
\begin{equation}
\begin{aligned}
\label{OPEsing}
X(z)X(0)\sim &\frac{(\cdots)}{z^2}\,,\qquad
   \xi(z)X(0)\sim\frac{(\cdots)}{z^2}\,,\\
Y(z)Y(0)\sim &\frac{(\cdots)}{z^2}\,,\qquad
   P(z)Y(0)\sim\frac{(\cdots)}{z^2}\,.
\end{aligned}
\end{equation}
If locality is not important, one can plug several
operators at different values of $z$ and get singularity-free multi-picture
changing operators in this way. Otherwise, the singular parts of the OPE's
can be simply removed, since they correspond to $Q$-exact terms.

With the understanding of picture changing operators one can define
more ways for representing the cohomology problem in the large
Hilbert space~\cite{Berkovits:2001us}. In these new representations, the
physical space could reside in the small Hilbert space and not in its $\xi_0$
copy as before. Hence, the ghost and picture numbers are the
same as for the $Q$ cohomology. The cohomology operator that one should use
is $\tQ$ of~(\ref{Qtilde}).

Let $V$ be a closed state with respect to $\tQ$, i.e.,
\begin{equation}
\label{tQV0}
\tQ V=0\,,
\end{equation}
and let it have a given picture number $p$.
Then, since $Q V$ and $\eta_0 V$ have different picture numbers,~(\ref{tQV0})
implies that,
\begin{equation}
Q V=\eta_0 V=0\,,
\end{equation}
i.e., $V$ lives in the small Hilbert space and is closed. Suppose we add to
$V$ a term of the form $\tQ \La$, subject to the constraint that the result
still has picture number $p$. Such a $\La$ can be decomposed to picture
numbers $p$ and $p+1$ plus a term $\La_{triv}$ obeying,
\begin{equation}
Q\La_{triv}=\eta_0\La_{triv}=0\,.
\end{equation}
Such a term does not contribute to $\delta V$ and so can be discarded.
The requirement that $\delta V$ has picture number $p$ implies,
\begin{equation}
Q \La_{p+1}=0\,,\qquad \eta_0 \La_p=0\,.
\end{equation}
The triviality of $Q$ and $\eta_0$ in the large Hilbert space then implies,
\begin{equation}
\La_{p+1}=Q \La^Q_{p+1}\,,\qquad \La_p=\eta_0 \La^\eta_{p+1}\,.
\end{equation}
All in all we can write,
\begin{equation}
\delta V=\tQ \La=(Q-\eta_0)\big(Q \La^Q_{p+1}+\eta_0 \La^\eta_{p+1}\big)=
  Q\eta_0(\La^Q_{p+1}+\La^\eta_{p+1}\big)\,,
\end{equation}
that is, two states are identified if they differ by a state, which is
$Q$ exact in the small Hilbert space. Given such a variation,
\begin{equation}
\delta V=Q\La\,,
\end{equation}
with $\La$ in the small Hilbert space, one can choose
\begin{equation}
\La_p=\La\,,\qquad 	\La_{p+1}=0\,,
\end{equation}
and get this variation as a variation by a $\tQ$ exact term in the large
Hilbert space.
Hence, the cohomology of $Q$ in the small Hilbert space is the same as that
of $\tQ$ in the large Hilbert space at the same picture and ghost
numbers, as stated.

Unlike the case of the usual cohomology problem, the cohomology does not
change upon relaxing the fixed picture condition. Assume that the
state $V$ carries a picture number bounded between
$p_{min}$ and $p_{max}$. We write,
\begin{equation}
V=\sum_{p=p_{min}}^{p_{max}} V_p\,.
\end{equation}
Consider the gauge transformation generated by\footnote{Having string field
theory in mind, we refer to a change of the representative of a given
cohomology as a gauge transformation.},
\begin{equation}
\label{gaugeUp}
\La=\xi(z)V_{p_{min}}\,.
\end{equation}
It induces the variation,
\begin{equation}
\delta V=(Q-\eta_0)\xi(z)V_{p_{min}}=(X-\xi Q) V_{p_{min}}-V_{p_{min}}
  +\xi\eta_0 V_{p_{min}}\,.
\end{equation}
The first term has picture number $p_{min}+1$, the second term eliminates the
original $V_{p_{min}}$, while the last term drops out, since it is the
lowest picture component of~(\ref{tQV0}).
Thus, this gauge transformation removes the lowest picture of the state.
If one considers instead the gauge transformation generated by
\begin{equation}
\label{gaugeDown}
\La=-P(z)V_{p_{max}}\,,
\end{equation}
one removes the highest picture component, since now
\begin{equation}
\delta V=(Y-P\eta_0) V_{p_{max}}-V_{p_{max}}\,.
\end{equation}

Using these transformations, one can reduce the picture number range of any
state to a single arbitrary picture number,
\begin{equation}
p_{min}\leq p \leq p_{max}\,.
\end{equation}
For this case, we have already shown that the $\tQ$ cohomology is equivalent
to the $Q$ cohomology.
Moreover, the equivalence to the $Q$ cohomology is independent on the choice
of the final $p$. To show that, consider the one before last stage in the
sequence of gauge transformations, where two non-trivial picture numbers
$p$ and $p+1$ are left. Using the gauge transformations~(\ref{gaugeUp})
or~(\ref{gaugeDown}), we can get to either of,
\begin{subequations}
\begin{align}
V_p+V_{p+1}& \rightarrow V_p+YV_{p+1}-P\eta_0 V_{p+1}\,,\\
V_p+V_{p+1}& \rightarrow X V_p+V_{p+1}-\xi Q V_p\,.
\end{align}
\end{subequations}
For the first two terms in the r.h.s of the two equations, it is obvious that
they are picture changed versions of the same state.
For the last term it follows from~(\ref{PYxiX}) and
\begin{equation}
QV_p-\eta_0 V_{p+1}=0\,,
\end{equation}
which is the $p$-picture component of~(\ref{tQV0}) for this case.
We can now conclude that the cohomology of $\tQ$ over the space of states
with picture number, which is arbitrarily bounded from both sides, is
canonically isomorphic to that of $Q$ at any fixed picture number.

Assume now that the picture number is unbounded.
By indefinitely repeating the procedure defined above, one can send the
picture to arbitrarily high or low values. Hence, the state component
at any given picture eventually will be zero. Explicitly, consider,
the multi-picture changing operators $X_n$ and $Y_{-n}$, for $n\geq 0$.
These operators are the regularized versions of the powers of $X$ and $Y$,
which are otherwise divergent and will be properly defined in the next
section. In particular,
\begin{equation}
X_1=X\,,\qquad Y_{-1}=Y\,,\qquad X_0=Y_0=1\,.
\end{equation}
Among other properties, theses operators satisfy the relations,
\begin{align}
\label{XnYnProp}
Q X_n&=Q Y_{-n}=\eta_0 X_n=\eta_0 Y_{-n}=0\,,\\
\label{XYnYXn}
X Y_{-n}&\sim Y_{-(n-1)}\,,\qquad Y X_n\sim X_{n-1}\,.
\end{align}
Using these operators we can show that any $\tQ$-closed state is also exact,
since the following contracting homotopy operators exist,
\begin{equation}
\label{QtContractHomo}
\La_-=\xi\sum_{n=1}^\infty Y_{-n}\,,\qquad
 \La_+=-P\sum_{n=1}^\infty X_n\,.
\end{equation}

One might mistakenly conclude that the cohomology of $\tQ$ over the space of
states with unbounded picture number is trivial. However, there is always
more than one way to define spaces using infinite sums of objects living is
some constituent spaces.
In the case at hand, we assumed that convergence in the all-picture space
is defined ``point-wise'', i.e., as the union of the limits at every picture.
There are certainly other ways to define limits. Let us use an analogy
with the following sequence of vectors,
$(1,0,...),\ (1,1,0,...),\ (1,1,1,0,...),...$
The point-wise limit of this sequence exists and equals $(1,1,1,...)$.
However, its limit in the $L_1$ norm, for instance, does not exist, since
the norm of the ``would-be limit'' diverges.

One could look for the analogy of the above for the case of vertex operators.
Any physical vertex operator can be represented in all possible pictures.
Moreover, there would also be many exact terms that could be added to it.
We can construct a $\mathbb Z$-sequence for any infinite sum of
representatives of the physical vertex operator in the following way.
We identify the location along the vector with the picture number. We choose
a representation for the vertex operator at some given picture number and
associate it with the vector $(...,0,0,1,0,0,...)$. Next, we identify all the
exact states with the zero vector and let the operators $X$ and $Y$ shift
the vectors to the left and to the right. These rules establish a unique
assignment of vectors. We now want to constrain the space of vectors by
imposing some norm on it. The $L_n$ norms are probably not what we are
after, since it is natural to expect that the norm commutes with the
operations $X$ and $Y$.
Instead, consider the absolute value of the sum of all
entires, provided it is well defined regardless of the summation order. This
is only a semi-norm, since there are many elements whose ``norm'' is zero,
i.e., all the exact states and all the states that are represented by vectors
with entries that sum up to zero.
Nonetheless, the semi-norm is enough for defining the space
that we need. One can, as usual, divide this space by the trivial space and
obtain a genuine norm. The resulting normed space would be one-dimensional
and it would correspond to the cohomology problem of this vertex.

We do not know how to generalize this construction to cover the whole
space of string fields, since we do not know how should the relative
normalization of different physical vertices work, nor do we know what to do
with non-closed, i.e., off shell states.
This is, however, exactly the usual problem with the definition of the space
of string fields, which cannot be accomplished due to a lack of a natural
norm. We conclude, that in this respect the democratic theory is not better,
but also not worse than any other string field theory.
We would also like to point out, that the correct definition of the
space of string fields would probably differ from the one obtained
using the semi-norm presented above, even when restricted to a given
physical vertex, due to non-linearities. Specifically, we see below that the
gauge transformation associated with picture changing should be modified in
the case of an interacting theory.

We saw that depending on the exact definition of the space of string fields
with unbounded picture number, $\tQ$ has a trivial
cohomology, or the same cohomology as for the bounded case.
Presumably it might also be possible to get other results for its cohomology.
However, since we do not have a complete definition for the space of string
field, we follow the ``standard'' practice in string field theoretical
research and ignore this problem, implicitly assuming that this space is
somehow defined in a proper way.

\section{Constructing the theory}
\label{sec:SSFT}

Here, we want to define an RNS string field theory, which generalizes the
$\tQ$-description of vertex operators. It turns out that a single or
a bounded range of picture numbers are inadequate choices. Hence, the theory
will generalize the case where all picture numbers are allowed.
We say that the theory is defined in the ``democratic picture'', since
within this construction, all string fields, regardless of picture numbers,
have equal opportunity to influence the physics\footnote{The idea that in
some cases all pictures should contribute is not new~\cite{Callan:1988wz}.
However, we suggest that it ``might be a feature, not a bug''.}.
The construction of the theory is almost straightforward. The only subtlety
is that a non-trivial insertion is required. We start this section by
deriving the form of the theory in~\ref{sec:SSFTform} and devote the major
part of the section to the derivation and to the study of the insertion
in~\ref{sec:SSFTInsertion}.

\subsection{Constructing the form of the theory}
\label{sec:SSFTform}

Let us start by writing the free action for the NS sector.
The equation of motion we are after is
\begin{equation}
\label{freeEOM}
\tQ\Psi=0\,,
\end{equation}
and the action should be invariant under the gauge transformation,
\begin{equation}
\label{LinSFTgauge}
\delta \Psi=\tQ \La\,.
\end{equation}
From the discussion of the previous section we infer that $\Psi$ lives in
the large Hilbert space, has ghost number one and its picture number is
either bounded, or unbounded with some (unknown) restriction on the
behaviour of $\Psi$'s components as a function of the picture number.
It is natural to use $\Psi \tQ \Psi$ for the construction of the free action.
Since the string field lives in the large Hilbert space, we have to use the
large Hilbert space CFT expectation value for the integration over the space
of string fields. We can also consider some linear operation to be performed
before the integration. Hence, the free action should be of the form,
\begin{equation}
\label{linAction}
S_{free}=-\frac{1}{2}\oint\cO \big(\Psi \tQ \Psi\big)\,,
\end{equation}
where the action has been written with a canonical normalization despite the
fact that we haven't specified $\cO$ yet.

From the experience we have with other theories, we know that the equation of
motion and gauge symmetry can be naturally extended to the non-linear level
by writing,
\begin{equation}
\label{Action}
S=-\oint \cO\Big( \frac{1}{2}\Psi \tQ \Psi+\frac{1}{3}\Psi^3\Big).
\end{equation}
The infinitesimal gauge symmetry related to this action is,
\begin{equation}
\label{gauge}
\delta \Psi=\tQ\La+[\Psi,\La]\,,
\end{equation}
and its finite form is,
\begin{equation}
\Psi\rightarrow e^{-\La}\big(\tQ+\Psi\big)e^\La\,.
\end{equation}
An important consequence of the above is that if the original picture number
is allowed to be non-zero, the picture number cannot be bounded and we are
led to consider the space of (properly restricted) string fields with
arbitrary picture number.

\subsection{Constructing the mid-point insertion}
\label{sec:SSFTInsertion}

One might wonder whether a non-trivial $\cO$ is really needed.
Let us give two arguments in favor of a non-trivial $\cO$:
\begin{itemize}
\item The ghost number of $\Psi \tQ \Psi$ equals three, but the integration
      we are using is the large Hilbert space integration, which is
      non-vanishing only for ghost number two string fields. Hence, the
      action would be identically zero without an $\cO$ insertion and it
      would not imply the desired equation of motion.
\item The large Hilbert space integral picks up the copy of the small
      Hilbert space that is multiplied by $\xi_0$.
      Nevertheless, as we showed in the previous section, it is the small
      Hilbert space without this insertion that carried the physical
      information when $\tQ$ is used.
\end{itemize}
The only consistent form for $\cO$ is that of a mid-point insertion of a
zero-weight conformal primary\footnote{We refer, again, to section 2
of~\cite{Kroyter:2009zj}, for discussion on this subject. Also, note that
there are two ``mid-points'', due to the doubling-trick. One may
hope at this stage that both $z=\pm i$ will be proven to be equivalent or
that some specific superposition of the two will be forced on us by
requiring reality of the action, or by some other principle.}. The remarks
above suggest that this insertion should contain the $\xi$-field.
Another immediate restriction on
$\cO$ is that it has to commute with the kinetic operator, since otherwise
the equations of motion and gauge transformations will not work out
correctly,
\begin{equation}
\label{QcO0}
\tQ\cO=0\,.
\end{equation}
Decomposing the mid-point insertion with respect to the picture number,
\begin{equation}
\cO=\sum_{n \in {\mathbb Z}}\cO_n\,,
\end{equation}
turns the relation~(\ref{QcO0}) into a recursion relation for the
$\cO_n$'s,
\begin{equation}
\label{recursion}
Q\cO_n=\eta_0\cO_{n+1}\,.
\end{equation}

In order to be able to use the recursion relations, an initial condition
is also needed.
For finding an appropriate initial condition we invoke the ``correspondence
principle''. Let us fix the picture-related gauge symmetry by restricting
the string field to carry zero picture number and to live in the small
Hilbert space. This partial gauge fixing reduces the action to that of the
modified theory, provided that we choose,
\begin{equation}
\cO_{-1}=\xi Y_{-2}=\cY\,,
\end{equation}
where $\cY$ is given by~(\ref{YetaPot}).
This choice of $\cO_{-1}$ implies that all the classical solutions of the
modified theory~\cite{Erler:2007xt,Aref'eva:2008ad,Fuchs:2008zx}
are also solutions, with the same action and cohomology, of the new theory.
Moreover, it is also possible to generalize the construction of boundary
states~\cite{Kiermaier:2008qu} to the modified theory~\cite{Kroyter:2009bg}
and thus, also to our case.
All that gives much credibility to the construction.

Substituting $\cO_{-1}$ into the recursion relation~(\ref{recursion})
immediately leads to,
\begin{subequations}
\label{UniqueO01}
\begin{align}
\cO_0=& \xi Y=P\,,\\
\cO_1=& \xi\,.
\end{align}
\end{subequations}
We see that in the three examples above, the insertion is given by the
product of $\xi$ and the picture changing operators $Y_{-2},Y,1$. This
is the case since the picture changing operators themselves are invariant
under $Q$ and $\eta$, while upon acting on $\xi$, $Q$ produces the picture
changing operator $X$.
This state of affairs cannot continue to $|p|>1$ picture
numbers without a modification, due to the OPE singularities~(\ref{OPEsing}).
A straightforward resolution is
to consider only the non-singular part\footnote{Note, that the fact that
our insertion includes all possible pictures is important, since it
enables a non-trivial value for the action for string fields of
arbitrary picture number.},
\begin{equation}
\label{On}
\cO_n= \left\{\begin{array}{ll}
  {\displaystyle \oint_w \frac{dz}{2\pi i} \frac{\xi(z)X_{n-1}(w)}{z-w}}&
     n>1\\ \\
  {\displaystyle \oint_w \frac{dz}{2\pi i} \frac{P(z)Y_n(w)}{z-w}}
    \qquad \qquad & n<-1\ \,,\\
 \end{array}\right.
\end{equation}
where $X_p$ and $Y_p$ are the multi-picture-changing operators.

One might wonder whether this construction is reliable, e.g., whether
multi-picture-changing operators exist at all and whether they are unique
in some sense. In fact they are. To explain this claim, let us recall some
more known facts about the RNS string~\cite{Horowitz:1988ip,Lian:1989cy}:
\begin{itemize}
\item The BRST operator $Q$ commutes with the ghost and picture generators.
      Hence, the cohomology can be decomposed to definite ghost and picture
      numbers.
\item The cohomologies at the same ghost numbers and different picture
      numbers are isomorphic.
\item For non-zero momentum the cohomology is concentrated in ghost numbers
      one and two, which are in fact (Poinear\`e) dual.
\item Unique non-trivial elements exist in the cohomologies at zero
      momentum at ghost numbers zero and three (and all picture numbers).
\end{itemize}
These facts imply that picture changing operators exist for all
integer picture numbers. We can also deduce that these
multi-picture-changing operators are unique, up to the addition of $Q$-exact
terms. We denote these operators by $X_n$, that is we define,
\begin{equation}
\label{XnDef}
X_n\equiv \left\{
\begin{array}{cc}
X_n &\qquad n>0\\
1 &\qquad n=0\\
Y_n &\qquad n<0
\end{array}
\right.\,.
\end{equation}
The OPE $X_n X_m$ might generally contain singular terms. These singular
terms, however, must be $Q$-exact, as can be seen from plugging the OPE into
a general expectation value and using the observations above.
For similar reasons, the regular term must equal $X_{n+m}$,
\begin{equation}
\label{XnXmXnm}
X_n X_m = X_{n+m}+Q\mbox{-exact}\,.
\end{equation}
Even more explicitly we may write,
\begin{equation}
\label{XnDef2}
X_n(w)\equiv \left\{\begin{array}{ll}
  {\displaystyle \oint_w \frac{dz}{2\pi i} \frac{X_{n-1}(z)X(w)}{z-w}}&
     n>1\\ \\
  {\displaystyle \oint_w \frac{dz}{2\pi i} \frac{Y_{n+1}(z)Y(w)}{z-w}}
    \qquad \qquad & n<-1\\
 \end{array}\right.\,.
\end{equation}

With the definitions~(\ref{On}),~(\ref{XnDef}) and~(\ref{XnDef2}),
the insertion we propose can be schematically written as,
\begin{equation}
\label{NonPrimO}
\cO\simeq \xi\sum_{n=-\infty}^\infty X_n\,.
\end{equation}
It is amusing to notice that if we use~(\ref{XnXmXnm}) for replacing $X_n$
by $X^n$, forgetting for the moment about the ($Q$-exact) OPE divergences
and about convergence radius issues, we can write,
\begin{equation}
\cO\simeq \xi\Big(\frac{1}{1-X}+\frac{1}{1-X^{-1}}-1\Big)=0\,.
\end{equation}
A more relevant and more accurate observation is that, in the
language of~\cite{Berkovits:1999in,Berkovits:1994vy}, our insertion is the
``unique $U(1)$-neutral extension of the vertex operator 1''.
This is a very natural insertion for a democratic-picture formalism.
While we do not think that this is of a particular importance, we
note that the proposed mid-point insertion probably does not have
a non-trivial kernel. The existence of a regular string field whose OPE with
$\cO$ is zero at all the picture numbers, seems to us very unlikely.

So far we were ignoring the fact that $\cO$ has to be a primary field. We
would now like to prove that a primary solution to the recursion relation
exists. As a bonus, we will also prove that all the picture changing
operators have primary representatives.
The proof goes by induction for $p>1$ and for
$p<0$. The respective initial conditions are $\cO_1=\xi$ and $\cO_0=P$,
which are completely fixed by their quantum numbers and are indeed
primaries. Acting on a zero-weight primary field with either $Q$ or $\eta_0$
gives back a primary. Hence, we can assume that the multi-picture changing
operators are primaries and we only have to prove that the $\cO_p$'s are also
primaries. Given that $\cO_p$ is primary for some $p>0$, it follows
from~(\ref{On}) and~(\ref{XnDef2}) that $X_p=Q\cO_p$ is also primary. The
$\cO_{p+1}$ defined by~(\ref{On}) is not a primary. In fact, for $n\geq 0$
we can write,
\begin{equation}
\label{LnO}
L_n \cO_{p+1}=\oint \frac{dz}{2\pi i}z^{n+1} \frac{dw}{2\pi i}
      T(z)\frac{\xi(w)}{w}X_p(0)=
   \oint \frac{dw}{2\pi i}w^{n} \partial \xi (w)X_p(0)=
   -n \xi_n X_p (0)\,.
\end{equation}
Here, the first equality is a mere substitution. In the second equality
we used the fact that $X_p$ is a primary. Integration by parts leads to
the final result. We note, that while $\cO_{p+1}$ is not a primary, the
problematic terms introduced by acting on it with the Virasoro operators
all lie in the small Hilbert space. Hence, we could try
to cancel them by adding to $\cO_{p+1}$ a term, which resides in the small
Hilbert space. Such a term would not lead to a change of $X_p$, so the
recursion relations at lower orders are intact. Let us prove that such
a term exists.

Let $\cS$ be the set of non-zero elements obtained from repeatedly acting
on $\cO_{p+1}$ with positive Virasoro operators. The level of this set is
bounded by some integer $N$.
Define $\cS_N$ to be the set of all possible level-$N$ combinations of
Virasoro operators acting on $\cO_{p+1}$,
\begin{equation}
\cS_N=\{L_N \cO_{p+1},L_{N-1}L_1 \cO_{p+1},\ldots\}\,.
\end{equation}
Some of the elements of $\cS_N$ are actually zero, but let us consider the
more general case, in which the $\cO_{p+1}$ operator is replaced by an
arbitrary (non-primary) zero-weight operator,
whose maximal level under the action of multiple Virasoro operators is $N$.
The size of $\cS_N$ is
$P(N)$, the number of partitions of $N$ into positive integers. Let $\cC_N$
be the set of all elements that can be formed by acting on $\cS_N$ with an
arbitrary number of negative Virasoro operators of total level $-N$.
The set $\cC_N$ contains $P(N)^2$, a-priori independent, zero-weight,
small Hilbert space elements. 
These elements are annihilated by the action of Virasoro operators of
total level greater than $N$, but not by operators of level $N$ or lower.
Consider now
\begin{equation}
\cO_{p+1}^{(N)}=\cO_{p+1}+\sum_{k=1}^{P(N)^2} \al_k C_k\,,
\end{equation}
where $C_k\in \cC_N$ and the $\al_k$'s are coefficients. Act on
$\hat \cO_{p+1}$ by all ($P(N)$) possible level N Virasoro operators
and require that the result vanishes.
This gives $P(N)$ equations, where each equation contains the $P(N)$
elements of $\cS_N$, which should vanish separately. All in all, we get
$P(N)^2$ linear non-homogeneous equations in $P(N)^2$ variables. In fact,
it is easy to see that the matrix describing these $P(N)^2$ equations
factorizes into $P(N)$ identical blocks of size $P(N)\times P(N)$.
Each block is nothing but the level-N Kac-matrix with $c=0$ and $h=-N$.
The Kac determinant at level $N$ equals up to a constant to the product of
terms of the form,
\begin{equation}
P_{r,s}=h-h_{r,s}\,,\qquad
h_{r,s}=\frac{c-1}{24}+F_{r,s}^2\,,
\end{equation}
with $F^2_{r,s}$ known, $c$-dependent positive constants.
In our case $h$ is a negative integer, while
\begin{equation}
h_{r,s}\geq -\frac{1}{24}\,.
\end{equation}
Hence, the Kac-determinant is non-zero and a solution exists and is
unique. The solution defines
$\cO_{p+1}^{(N)}$ as a zero-weight operator that differs from our
original $\cO_{p+1}$ only in the small Hilbert space. Acting on
$\cO_{p+1}^{(N)}$ with Virasoro operators of total level greater than
$N-1$ gives zero by construction. We can now repeat the
procedure, defining $\cO_{p+1}^{(N-1)}$ that removes the $N-1$ terms and so
on. The operator $\cO_{p+1}^{(1)}$ is then primary by construction.
The fact that many of the
elements of $\cC_N$ are zero does not modify any of the arguments
we made and for $p>1$ the proof is completed. The proof for $p<0$ is
almost identical. All that is needed is to replace $Q$ by $\eta_0$,
$\xi(z)$ by $P(z)$ and the small Hilbert space by the dual
small Hilbert space.

While our proof implies using $P(N)^2$ elements for eliminating $\cS_N$,
we can use the explicit form~(\ref{LnO}) in order to work in practice
with smaller sets. In fact, generalizing~(\ref{LnO}), we see that
all the elements of $\cS_N$ are equal up to a constant. Hence, we can
replace the $P(N)^2$ elements of $\cC_N$ by the $P(N)$ elements
obtained by acting on $\xi_N X_p(0)$ with total-level $-N$ Virasoro
operators. We now get a set of at most $P(N)$ equations with at most
$P(N)$ variables. The matrix of coefficient is again the Kac matrix,
which leads to a unique solution.
This is a simplified (but less general) version of the theorem above.

Let us now consider $\cO_2$ as an example. The theorem implies that
it is given by,
\begin{equation}
\cO_2=-c\xi\xi'+\xi e^\phi G_m
  +\Big(2 b \eta \xi \phi '+\eta  \xi  b'-2 b \xi  \eta '\Big)e^{2\phi}
  +\tilde \cO_2\,,
\end{equation}
where $\tilde \cO_2$ resides in the small Hilbert space. Here, we truncated
the expression one gets using~(\ref{On}) to its $\xi_0$ component, as
the rest of it lies in the small Hilbert space, which we can freely change.
Applying the Virasoro operators, we see that $N=2$ in this case and that
\begin{equation}
\label{O2CT}
-L_1^2\cO_2=L_2\cO_2=4b e^{2\phi}\,.
\end{equation}
Considering the two terms $L_{-1}^2 b e^{2\phi}$ and $L_{-2} b e^{2\phi}$
gives two equations, which we can solve. The solution, however, is quite
cumbersome when written explicitly. Moreover, while the solution is
universal by construction, it is not manifestly universal, e.g., for the
standard matter fields we get different
coefficients of $T_X$ and $T_\psi$ in various terms of the solution.
A way around it is to note that in order to prove the theorem, we do not
really have to use the full energy momentum tensor. All that is needed is to
have $T_{\xi\eta}$, as well as the part of the energy momentum tensor
associated with the fields that appear in~(\ref{O2CT}), i.e., we can work
with $T_g$ instead of with the total energy momentum tensor in the case at
hand. Working with $T_g$ gives again two equations. Solving the equations
leads to a $\tilde \cO_2$, which contains six, manifestly universal terms.
Actually, playing a bit with coefficients reveals an even simpler
solution,
\begin{equation}
\tilde \cO_2=-\Big(29b''+51b'\phi'
  +2b \phi '^2\Big)\frac{e^{2\phi}}{86}\,.
\end{equation}
This illustrates the fact that while a solution of the form used in the
proof above is unique, there are generally many other (universal) solutions.
It is interesting to note that although we explicitly removed only the
second order poles, we got a primary field, without actually going to the
next stage.

It is now straightforward to calculate $X_2$, which we did in a particular
background (one-dimensional linear-dilaton) for simplicity.
The resulting expression is a sum of 150 terms (in this background) and is
not particularly illuminating. This $X_2$ is primary by construction and
can be used to define $\cO_3$. Now, $N=6$ and terms with various
$\phi$-momenta appear. The $e^{4\phi}$-terms are found all the way to
$N=6$, while the $e^{3\phi}$-terms and $e^{2\phi}$-terms have $N=4$ and $N=3$
respectively. It is clear that all these terms would not be removed in a
single step.
We examined a 371-parameter universal ansatz, containing
the terms that are dictated by the theorem, and found a 94-parameter family
of solutions. In a relatively simple case a primary $\cO_3$ can be
written as a sum of 336 terms.

Having settled the issue of being primary, we now want to discuss the amount
of freedom in solving the recursion relation~(\ref{recursion}), without the
restriction to primary operators. Too much freedom
in choosing $\cO$ would lead to an embarrassment of riches, i.e., having
several, presumably inequivalent, theories. This state of affairs was
exactly one of the grounds for criticizing the modified theory. Then,
in~\cite{Kroyter:2009bg}, we showed that, at least classically, all these
theories are equivalent, regardless of the exact form of the insertion and
regardless of the distribution of the insertion among the two mid-points
($\pm i$).
The only thing we need in order to make the same assertion here, is to show
that the difference $\delta \cO$ of two candidate $\cO$ insertions is given
by a $\tQ$-exact term.
It is clear from~(\ref{QcO0}) that 
\begin{equation}
\tQ\delta \cO=0\,.	
\end{equation}
However, while $Q$ and $\eta_0$ are separately trivial in the large Hilbert
space, their difference, $\tQ$, is not. Hence, this equation is not
enough for our proof.

Nonetheless, the initial condition for the non-unique $\cO_{-1}$ fixes the
insertions $\cO_0$ and $\cO_1$ uniquely~(\ref{UniqueO01}), at least as long
as we are considering only chiral insertions. This can be seen simply by
examining all possible insertions with those quantum numbers.
Consider now $\cO_2$. The recursion relation~(\ref{recursion}) implies that
it is defined up to an $\eta_0$-closed term. Since $\eta_0$ is exact in the
large Hilbert space, we can write,
\begin{equation}
\delta \cO_2=-\eta_0 \Upsilon_3\,.
\end{equation}
This implies at the next picture number the identity,
\begin{equation}
\eta_0\delta \cO_3=Q\cO_2=\eta_0 Q \Upsilon_3\,,
\end{equation}
whose general solution is,
\begin{equation}
\delta \cO_3=Q \Upsilon_3 + \eta_0 \Upsilon_4\,.
\end{equation}
We see that the total effect of $\Upsilon_3$ is to induce a change in $\cO$,
\begin{equation}
\delta \cO=\tQ \Upsilon_3\,,
\end{equation}
and the analysis can then be repeated for $\Upsilon_4$ and so on.
A similar analysis can be applied for negative picture numbers.
We conclude that in general,
\begin{equation}
\delta \cO=\tQ \Upsilon\,.
\end{equation}
Hence, all the theories that are defined by solutions of the recursion
relation~(\ref{recursion}) with the initial conditions~(\ref{UniqueO01}),
are classically equivalent.
This also implies that, at least for the purpose of evaluating the action of
a classical solution, we do not have to find an explicit primary conformal
representative of $\cO$. The regular part of~(\ref{NonPrimO}) will do. Note,
however, that the two parts of the action would have to be evaluated in the
same coordinate system. This is not really a restriction, since for solutions
one simply evaluates,
\begin{equation}
S=-\oint \cO\Big( \frac{1}{2}\Psi \tQ \Psi+\frac{1}{3}\Psi(-\tQ \Psi)\Big)=
-\frac{1}{6}\oint \cO\Psi \tQ \Psi\,.
\end{equation}
Then, it is possible to perform the calculations using~(\ref{XnDef2})
rather than with the cumbersome explicit expressions of the conformal
primaries.

\section{The Ramond sector and general D-brane systems}
\label{sec:Ramond}

In order to complete this work, we have to incorporate the Ramond sector into
the formalism. This sector leads to various singularities within the modified
theory. Nonetheless, the construction of the action of the modified
theory, as well as of Witten's theory, relied on
correct principles. Hence, if we could manage to construct an action that
generalizes these constructions, without suffering from their problems, we
would know that we are on the right track. Thus, we search for an action
that reduces to,
\begin{equation}
S= -\int Y_{-2}\Big(\frac{1}{2}\Psi Q \Psi+\frac{1}{3}\Psi^3\Big)
   -\int Y\Big(\frac{1}{2}\al Q \al+\Psi \al^2\Big),
\end{equation}
when the NS and Ramond string fields live in the small Hilbert space and
carry picture numbers zero and $-\frac{1}{2}$ respectively.
From the inconsistency of the modified theory we can infer that the above
set of restrictions does not form a consistent set of gauge conditions.
We return to the issue of gauge fixing after the theory is constructed.

We know already that the first part of this action generalizes
to~(\ref{Action}). Then, applying
the same philosophy as before, we replace $Q$ with $\tQ$, the $Y$ insertion
with $\cO^R$ and the integration with integration in the large Hilbert space.
The correspondence principle implies that
\begin{equation}
\cO^R_0=P\,.
\end{equation}
Having $\tQ$ as the kinetic operator implies again the recursion
relations~(\ref{recursion}) and hence,
\begin{equation}
\cO^R=\cO^{NS}\,.
\end{equation}

It is straightforward to see that the resulting action can be simply written
in terms of a single string field and that it takes exactly the simple
form~(\ref{Action}), provided we redefine,
\begin{equation}
\Psi\rightarrow\Psi+\al\,.
\end{equation}
Thus, we unified the NS and Ramond string fields. The resulting string field
$\Psi$ includes all possible integer (NS) and half-integer (R) picture
numbers at ghost number one.
The full gauge symmetry of the action is~(\ref{gauge}), where the even gauge
string field $\La$ carries ghost number zero and an arbitrary integer (NS)
or half-integer (R) picture number. There are no picture
changing operators or other mid-point insertions in the definition of the
gauge symmetry. Hence, no singularities can emerge and the gauge symmetry is
well defined.

The inclusion of the various sectors (NS$\pm$, R$\pm$) of a general D-brane
configuration, described by some Chan-Paton factors, is
straightforward and should be done in the same way as
in~\cite{Arefeva:2002mb} (see also~\cite{Berkovits:2000hf}), i.e., the space
of string fields should be tensored with a matrix space representing the
Chan-Paton factors, as well as with the ``internal Chan-Paton space'' of
two by two matrices. The NS+ string field is tensored with an internal
Chan-Paton factor of $\sigma_3$ (granted also to $Q$), while the NS$-$
string field is tensored with $i\sigma_2$.
Since the Ramond string fields are added to the NS+ string field in our
formulation, they should also be tensored with $\sigma_3$. A single
Chan-Paton entry cannot contain both R sectors. Hence, there is no problem in
assigning the same factor to both R$\pm$. This assignment
is also consistent with our discussion on this subject
in~\cite{Kroyter:2009zi}. The fact that $Q$ gets a $\sigma_3$ factor implies
that $\eta_0$ gets the same factor (again, in accordance with the discussion
in~\cite{Fuchs:2008zx,Kroyter:2009zi}). This implies that $\xi$ and $P$
insertions should also be tensored with $\sigma_3$, which suggests that the
whole $\cO$ insertion is to be tensored with this factor. This, in a sense,
clarifies the origin of the $\sigma_3$ insertion on $Y_2$ in the modified
theory: It is a remnant from integrating the $\xi$ insertion that has
to carry this factor, when going from the large Hilbert space to the small
one. The gauge string fields should all carry a unity factor, except the
NS$-$ gauge string field that carries a $\sigma_1$ factor.
With these assignments all the axioms needed continue to hold and all
sectors of an arbitrary (not necessarily BPS) D-brane system can be
uniformly and covariantly described.

\section{Field--antifield formulation and gauge fixing}
\label{sec:BV}

The gauge symmetry of our theory is infinitely reducible and closes only
``using the equations of motion'', i.e., only up to trivial gauge
transformations. This state of affairs calls for the use of the
(covariant) field--antifield (BV)
formalism~\cite{ZinnJustin:1974mc,Batalin:1981jr,Batalin:1984jr,
Batalin:1984ss,Voronov:1982cp}
(see~\cite{Henneaux:1989jq,Henneaux:1992ig,Gomis:1994he} for reviews).
This formalism replaces the gauge symmetry by a BRST symmetry, which can
be fixed at a later stage using a ``gauge--fixing fermion''.
Generally speaking, the BV construction is  nothing but trivial.
Luckily, the algebraic structure at hand is identical to that of the
bosonic theory. The BV formulation of this theory was studied by Thorn and
by Bochicchio~\cite{Thorn:1986qj,
Bochicchio:1986bd,Bochicchio:1986zj,Thorn:1988hm}.
All what we have to do then, is to use the following substitutions,
\begin{equation}
Q_{bos} \rightarrow \tQ=Q_{RNS}-\eta_0\,,\qquad
 \int_{bos} \rightarrow \oint \cO\,,\qquad
  \Psi_{cl,bos} \rightarrow \Psi_{cl,RNS}\,,
\end{equation}
where as already explained, the classical RNS field $\Psi_{cl,RNS}$
carries ghost number one and all (integer and half-integer) picture numbers.
Mimicking the construction of~\cite{Thorn:1986qj,
Bochicchio:1986bd,Bochicchio:1986zj,Thorn:1988hm} for the case at hand is
straightforward, due to the identical algebraic structure (the properties of
$\tQ$ and $\oint \cO$ as well as the form of the redundant gauge
symmetry). The construction leads (before gauge fixing) to an action
identical in form to~(\ref{Action}), only with the string field
$\Psi_{cl,RNS}$ replaced by the string field $\Psi_{BV,RNS}$, which contains
all possible picture and ghost numbers, as well as both (NS and R) sectors
of the theory.
This is very satisfactory from an aesthetic point of view, since now the
theory is defined by a string field that uses the whole ``Hilbert space''.
Moreover, for the case of a general D-brane system, the string field
lives in the maximal space (in terms of sectors and ghost and picture
numbers) consistent with the system.

One possible subtlety with this construction is the implicit assumption that
the integration measure that we use induces a non-degenerate inner product
in the space of string fields. While the analogous assertions in some other
cases are pretty much CFT axioms, our case
might depend on the definition of the space of string fields. The reason
for the difference is that, in theories with no explicit insertions, the
space is naturally decomposed into subspaces with fixed ghost and picture
numbers and with fixed conformal weights.
Each such space is dual to another such space with respect to the
inner product. Hence, the string field can be decomposed into a direct sum of
spaces and within each of these spaces it can be further decomposed into
a finite sum of component fields. Each component field has as its anti-field
another component field that lives in its dual space. All that implies that
the BV theory can be equally well formulated in
terms of the component fields and in terms of string fields.

When mid-point insertions are included, they tend to couple fields with
various conformal weights. Moreover,
in our case, there is a decomposition only with respect
to the ghost number, while all picture numbers are coupled, due to the
presence of the $\cO$ insertion, which includes all integer picture numbers.
The inner products between the elements of two dual spaces at ghost numbers
$g$ and $3-g$ can be collected into an infinite matrix. The question of
degeneracy then depends on the space of allowed vectors and dual vectors
(ghost number $g$ and $3-g$ string fields respectively). One observation is
that this matrix contains
infinitely many infinite size blocks, of constant entries. One could
fear that these blocks, originating from sets of physically equivalent
vertex operators, imply that the measure is degenerate. This is not
necessarily the case, since the elements generating such a block
couple also to other, in particularly off-shell, states.

We would like to stress that it would be wrong to try and ``invert''
$\cO$ treating it separately from the measure, as it is often done
with the $Y_{-2}$ of the modified theory. Particularly, the inverse
might not be defined over the correct space of string fields, i.e.,
the formal object ``$\cO^{-1} \Psi$'' would probably not be part of the
space of string fields for any non-zero $\Psi$.

We believe that the subtleties related to the definition of the space of
string fields have a resolution, the BV construction is reliable and
can be used for gauge fixing. To that end, a gauge-fixing
fermion should be introduced.
The gauge fixing fermion is an odd functional of the string fields, with
(second quantized) ghost number $-1$. It depends on some auxiliary fields,
e.g., non-minimal sets of variables that serve as Lagrangian multipliers.
The set of non-minimal fields needed for our case is
known~\cite{Gomis:1994he}. However, there are many ways for constructing
gauge fixing fermions, not all of which are admissible. Moreover, it is not
clear whether the subtleties with the quantum master
equation of the bosonic theory persist in our theory.
Even if it would be possible to show that the quantum master equation holds
in our case, the construction of an admissible gauge fixing fermion would
probably still be non-trivial. In calculating RNS loop scattering amplitudes
subtleties appear related to the measure on supermoduli
spaces~\cite{D'Hoker:2002gw,Morozov:2008wz,DuninBarkowski:2009ej}. It might
be too optimistic to expect that
these issues can be avoided by using string field theory for the
evaluation of amplitudes. On the other hand, if one could show that
superstring field theory is consistent at the quantum level, this could be
an alternative definition for the superstring, not relying on the subtleties
of supermoduli spaces!
We leave the resolution of these important questions to future work.

A somewhat naive alternative to the discussions above would be to simply
enforce some auxiliary conditions that suppose to remove
the gauge-related redundancy. For example, when the theory
is restricted to the NS sector, it is possible to constrain the string field
to carry zero picture number and to live in the small Hilbert space. The
action then reduces to that of~\cite{Preitschopf:1989fc,Arefeva:1989cp},
where a further gauge fixing is needed, e.g., Siegel gauge, Schnabl gauge, a
linear $b$-gauge~\cite{Kiermaier:2007jg}, or an
$a$-gauge~\cite{Asano:2006hk,Asano:2008iu}.
Trying to restrict the NS sector to other picture
numbers seem to lead to inconsistent results. At picture number $-1$, one
might argue that Witten's theory is obtained, while at other picture numbers
other inconsistent theories seem to emerge\footnote{After the first version
of this paper appeared, it was discovered in~\cite{Kroyter:2010rk} that
gauge fixing the theory to picture number $-1$ does not lead to Witten's
theory, but to another, classically consistent, theory, which is the $\Ztwo$
dual of the modified theory in the sense
of~\cite{Berkovits:1994vy,Berkovits:1996bf}.
Moreover, it was found there that the democratic theory can be reduced using
another gauge fixing to the non-polynomial theory and that this partial gauge
fixing can also be extended to the Ramond sector. Finally,
it is also argued there that gauge fixing at other fixed picture numbers is
probably inconsistent. These new results give strong evidence in favour of
the democratic theory.}. It would be very interesting to consider universal
gauge fixings that do not concentrate at fixed picture numbers. We leave
this interesting project to future work.

\section{Supersymmetry}
\label{sec:SUSY}

Up to this point, our discussion was universal, i.e., it did not
depend in any way on the BCFT used in the definition of the theory.
Now, however, we would like to
study supersymmetry within our theory, which is
background dependent. Hence, we specify that we work in the standard
ten-dimensional RNS flat space, i.e., the matter system is composed of ten
world-sheet scalars $X^\mu$ and ten world-sheet fermions $\psi^\mu$.

The space-time supersymmetry generators of the RNS formalism
carry half-integer picture numbers. In a fixed picture number theory this
implies that picture changing operators should be appended to the definition
of the supersymmetry transformation. For consistency, these operators have
to be inserted at the string mid-point, which leads to
singularities and probably takes the string field outside its domain
of definition\footnote{We stress again that general mid-point operator
insertion {\it on the string field} might lead to singularities. In order
to avoid these potential problems one has to restrict somehow the space
of string fields, such that potentially harmful mid-point insertions
would not be allowed. This in turn implies that mid-point operator
insertions {\it in the action}, as we consider here, are harmless.}.
Working at an unrestricted picture-number space, as we do here, potentially
avoids this problem.

In previous cubic formulations~\cite{Witten:1986qs}, supersymmetry was
generated by the zero momentum, integrated, $-\frac{1}{2}$-picture, fermion
vertex. We may consider the same generator in our theory,
\begin{equation}
\label{WittenSUSY}
\delta_{SUSY}\Psi=\ep^\al Q_\al \Psi\equiv Q^\ep\Psi\,,\qquad
Q_\al=\oint \frac{dz}{2\pi i}\, e^{-\frac{\phi}{2}}S_\al(z)\,.
\end{equation}
Here, $S_\al$ is the spin field, which is responsible for exchanging
the NS and Ramond sectors while $\ep^\al$ are odd parameters.
Using the integrated vertex seems to be the better option, since the
unintegrated vertex should be inserted at a given point. This point cannot
lie inside the local coordinate patch, since that might lead to
singularities from collisions with the state itself. It also cannot be
inserted outside the local coordinate patch, as that might take the string
field outside of its domain of definition, as well as to introduce
singularities from multiplication by other string fields.

The generator~(\ref{WittenSUSY}) would be a symmetry, provided that the
following three conditions hold~\cite{Witten:1986cc,Witten:1986qs}:
\begin{subequations}
\label{3conds}
\begin{enumerate}
\item $Q^\ep$ should be a derivation of the star product,
\begin{equation}
\label{FirstCond}
Q^\ep(AB)=Q^\ep A B+A Q^\ep B\,.
\end{equation}
\item $Q^\ep$ should be invariant under the kinetic operator,
\begin{equation}
\label{SecondCond}
[\tQ,Q^\ep]=0\,.
\end{equation}
\item $Q^\ep$ should leave the integration measure invariant,
\begin{equation}
\label{ThirdCond}
\oint \cO Q^\ep A=0\qquad \forall A\,.
\end{equation}
\end{enumerate}
\end{subequations}
The first condition holds, since $Q^\ep$ is an integral of a current.
The second condition holds, since the vertex defining $Q^\ep$ lives in the
small Hilbert space and is on-shell. The third condition, however, does not
hold, since the vertex has singular OPE's with many of the $\cO_n$'s.
This is hardly surprising, since these operators are essentially picture
changing operators, which should act non-trivially on an on-shell vertex
at any given picture.

While our supersymmetry generators fail to be symmetries off-shell, they
are symmetries on-shell. By that we do not mean that the action around
solutions is invariant under the linearized supersymmetry transformation:
The action is linearly invariant under any change of a solution by
definition. We mean that this linearized transformation naturally acts on
the space of solutions, i.e., it sends solutions to solutions,
\begin{equation}
\delta_{SUSY}(\tilde Q\Psi+\Psi^2)=\tQ Q^\ep \Psi
  + Q^\ep\Psi\Psi+\Psi Q^\ep \Psi
=Q^\ep(\tilde Q\Psi+\Psi^2)=0\,.
\end{equation}
Here used was made of the
properties~(\ref{FirstCond}) and~(\ref{SecondCond}).

Nonetheless, a genuine symmetry must be defined also off-shell\footnote{The
situation we have should not be confused with the common one of having a
symmetry algebra that ``closes only up to the use of the equations of
motion''. In the case at hand, we cannot even claim that we have a
symmetry.}. One possible direction towards defining supersymmetry off-shell
is to consider a superposition of supersymmetry generators at various picture
numbers,
\begin{equation}
Q_\al\stackrel{?}{=}
  \oint\frac{dz}{2\pi i} \sum_{p\in (\cz+\frac{1}{2})} k_p V^p_\al(z)\,,
\end{equation}
where $k_p$ are unknown coefficients. These coefficients are restricted by
the requirement that~(\ref{ThirdCond}) holds. One might think that this
restriction gives a set of recursion relations, similar to the ones that we
got for the $\cO_n$ insertions. Instead, one gets
an infinite set of equations, each one
including infinitely many summands. Each summand can include many different
operators that should all independently vanish by the choice of coefficients.
It is neither clear to us whether this system of equations has a
solution, nor how to construct this solution perturbatively, or otherwise.

A more promising approach can be based on the fact that (when integrated)
the fermion vertex is $\tQ$-closed.
Our experience with string field theory from the last few years
suggests that it might be useful to write it formally as if it were
exact~\cite{Okawa:2006vm,Fuchs:2007yy,Fuchs:2007gw}.
We propose to write the integrated fermion vertex as,
\begin{equation}
V_\al=-\tQ W_\al\,.
\end{equation}
Being loyal to the democratic paradigm, the above vertex should be allowed to
have an arbitrary picture and choosing different pictures should result in
gauge equivalent configurations. We would then write,
\begin{equation}
\label{QalQWTakeOne}
Q^\ep \Psi=\tQ (W^\ep) \Psi\,.
\end{equation}
The problem is that the string field is not necessarily
closed, so~(\ref{ThirdCond}) is still not obeyed, which is not surprising
since all we did was to rewrite the form of the vertex.
The following modification can be proposed in order to resolve this problem,
\begin{equation}
\label{QalQWPsi}
\delta_{SUSY} \Psi\stackrel{?}{=}\tQ (W^\ep \Psi)\,.
\end{equation}
Both~(\ref{QalQWTakeOne}) and~(\ref{QalQWPsi}) agree when restricted to
vertex operators and reduce in this case to the standard expression.
However, while~(\ref{ThirdCond}) holds
for~(\ref{QalQWPsi}),~(\ref{FirstCond}) no longer holds.
The natural way to resolve this problem is to notice that~(\ref{QalQWPsi})
is in fact the linearized form of an infinitesimal fermionic gauge
transformation\footnote{The idea of formally representing supersymmetry as a
gauge symmetry, albeit in a different way, appeared already
in~\cite{Qiu:1987dv}.}.
Hence, it is natural to add to the above a non-linear term and define,
\begin{equation}
\label{SUSYgauge}
\delta_{SUSY} \Psi=\tQ (W^\ep \Psi)+[\Psi,W^\ep \Psi]\,.
\end{equation}
The relations~(\ref{3conds}) are not obeyed now, but there is no reason for
them to be obeyed, since~(\ref{SUSYgauge}) is not a linear transformation of
the string field. It is nevertheless a symmetry, as it is a formal gauge
symmetry (recall that we use a formal gauge string field)\footnote{One might
have considered $W_\al=-\int_{-\infty}^\infty dz V_\al(z)\ket{1}$ as the
formal gauge string field, with $\ket{1}$ being the identity string field
and the integration is in the cylinder coordinates. The gauge string field is
manifestly formal due to the presence of the identity string field in its
definition and the resulting transformation is exactly~(\ref{WittenSUSY}).
Nonetheless, this choice is wrong, since the integration limits approach the
mid-point insertion, invalidating the arguments that show that a gauge
symmetry is a symmetry.}, with the (formal) gauge string field
\begin{equation}
\La_{SUSY}=W^\ep \Psi\,.
\end{equation}

The main potential problem with this proposal is to find an adequate $W^\ep$,
that is, to define it in such a way that while it is a formal string field,
the resulting transformation~(\ref{SUSYgauge}) defines a genuine
string field. Since $V_\al$ lives in the small Hilbert space, it is possible
to write,
\begin{equation}
\label{WalTakeOne}
W_\al\stackrel{?}{=}\xi V_\al\,.
\end{equation}
This, however, results in an addition to the vertex,
\begin{equation}
\delta V_\al=-Q W_\al\,,
\end{equation}
i.e., we add to the vertex an expression, which is roughly minus the same
vertex in a different picture. This is obviously wrong.
Indeed,~(\ref{WalTakeOne}) is a genuine
string field. Hence, the resulting transformation is a genuine gauge
transformation rather than a supersymmetry one. A way out might be to add
to $W_\al$ $\eta$-primitives of all higher pictures. Recall that up to
$Q$-exact (singular) terms, the recursion relation satisfied by the
integrated vertex operators is,
\begin{equation}
\label{Vrecurs}
V^{p+1}=Q(\xi V^p)+\partial(\xi \hat V^p)\,,
\end{equation}
where $V^p$ is the unintegrated vertex operator, defined by the relation,
\begin{equation}
\label{QVhatV}
QV^p=\partial \hat V^p\,.
\end{equation}
The second term in~(\ref{Vrecurs}) is needed for assuring that the resulting
vertex lives in the small Hilbert space and continues to
respect~(\ref{QVhatV}).
The recursion relation~(\ref{Vrecurs}) might suffer from OPE singularities.
However, as these singularities are exact, they can be safely removed from
the definition. Total derivative and $Q$-exact terms could then be added
to assure that the operator is primary.

Now, let us define,
\begin{equation}
W_\al^{p+1}=(-1)^{p-p_0} \xi V^p_\al\qquad p+1\geq p_0\,,
\end{equation}
and
\begin{equation}
W_\al=\sum_{p=p_0+1}^\infty W_\al^p\,.
\end{equation}
Here $p_0$ is an arbitrary starting point. 
With this definition we get
\begin{equation}
\tQ W_\al=-V_\al^{p_0}\,.
\end{equation}
Hence, at the linearized level, the transformation reduces to a supersymmetry
transformation.
Moreover, the components of $W_\al$ do not decrease in any way as a function
of picture number. Thus, it is indeed a formal gauge string field, as we
wanted to have.
The difference between starting with two different pictures, on the other
hand, is given by a genuine gauge transformation.

The resulting supersymmetry transformation takes the form,
\begin{equation}
\delta_{SUSY} \Psi=Q^\ep\Psi-W^\ep(\tQ\Psi+\Psi^2)\,.
\end{equation}
The second term vanishes on-shell and the transformation reduces to the
standard linear supersymmetry transformation. Off-shell the transformation
becomes non-linear with respect to the string field. This might have
been expected, since on the one hand string field theory is a non-linear
extension of the world-sheet formalism, while on the other hand
supersymmetry is non-linearly realised in many circumstances.

There is one potential obstacle for our interpretation: It seems that
the string field we obtain off-shell is not more genuine than the formal
gauge string field.
Indeed, our experience with formal gauge string fields suggests that
``higher order counter-terms'' should be added to the
gauge string field in order to obtain a legitimate physical string field.
We leave the problem of finding these terms and the related problem of
understanding the non-linear terms induced by the supersymmetry algebra in
the momentum transformation of off-shell string fields to future study.

\section{Conclusions}
\label{sec:conc}

In this work a new universal open RNS string field theory was presented.
The theory is cubic and includes a mid-point insertion in the action.
This mid-point insertion most probably carries only a trivial
kernel. More importantly, while we have an insertion in the action,
we allow for mid-point insertions neither in the gauge transformations nor in
the equations of motion. Thus, the theory does not suffer from singularities
due to collisions of mid-point insertions as was the case with the previous
formulations.

The new theory naturally and covariantly unifies the NS and Ramond string
fields and can be used to describe open strings on arbitrary D-brane
systems. Since it can be reduced to the modified theory, it supports
its solutions.
These observations give much credibility to the theory.

Nonetheless, there is more to be done. Specifically, one should devise a
gauge fixing of the theory, since it is imperative for the
construction of perturbation theory, e.g., for defining propagators.
While the first step towards that
end, i.e., the BV construction, was completed and led to an elegant
result, it is important to further study this highly non-trivial issue. The
understanding of the gauge fixing of our theory might lead not only to
further credibility to the theory, but also to new
ways of evaluating RNS scattering amplitudes, which avoid the problems with
the supermoduli spaces, at least for the scattering of open strings.
It is also important to complete the formulation of the off-shell
supersymmetry transformations of the theory.

Our construction uses primary multi-picture changing operators. The existence
of these operators was not known and was proven here. Moreover, we showed
that within a specific universal space spanned by total Virasoro operators
plus a non-trivial piece, these operators are unique. However, when one
considers more general, but still universal, spaces, these operators are
no longer unique. The understanding of multi-picture changing operators
achieved in this work is an important achievement by itself, since it
might be of use also for other approaches towards the RNS string.

We used the multi-picture changing operators in order to construct the
(primary) $\cO_n$ operators, which generalize $\xi$ and $P$ to other picture
numbers. The generalization is,
however, incomplete, in the sense that while $\xi$ and $P$ serve as
contracting homotopy operators for the commuting cohomology operators $Q$
and $\eta_0$, we do not know of currents that we can similarly associate
with the other $\cO_n$ operators. It would be interesting if we could
define currents $J_p(z)$, carrying ghost number one and picture number $p$,
such that the $\cO_{-p}$ would serve as the contracting homotopy operators
for their mutually commuting charges, i.e.,
\begin{subequations}
\begin{align}
J_p(z)J_{p'}(w)&\sim\ldots+\frac{\partial(..)}{z-w}\,,\\
T(z)J_p(w)&\sim\frac{J_p(w)}{(z-w)^2}+\frac{\partial J_p(w)}{z-w}\,,\\
J_p(z)\cO_{-p}(w)&\sim \frac{1}{z-w}\,.
\end{align}
\end{subequations}
The familiar cases are,
\begin{equation}
J_0(z)=J_B(z)\,,\qquad J_{-1}(z)=\eta(z)\,.
\end{equation}
We managed to find a candidate $J_{-2}$,
\begin{equation}
J_{-2}=\frac{1}{4}\big(3 c''-13 c\phi'^2+2(bcc'+c\xi'\eta+c T_m)\big)
        e^{-2\phi}\,.
\end{equation}
In fact, we even found a multi-parameter family of such candidates.
There might, however, be further restrictions on $J_{-2}$ coming from
consistency with the other $J_p$'s, or otherwise.
We believe that a better understanding of these currents (if they exist)
might shed light on the nature of the RNS string, as well as help in future
string field theoretical research.

Our construction is based on the use of $\tQ$, which was introduced by
Berkovits in~\cite{Berkovits:2001us}, for the purpose of relating the RNS
formalism and the pure-spinor one~\cite{Berkovits:2000fe,Berkovits:2002zk}.
Here, we used this operator for constructing an RNS string field theory. One
might ask whether it is possible to rewrite our theory using pure-spinor
variables. This is not
straightforward, since the mapping between the two formalisms is for
on-shell states only, while we should be looking for an off-shell mapping.
On the other hand, a pure-spinor string field theory already
exists~\cite{Berkovits:2005bt}. This theory is also cubic and it also uses
a mid-point insertion whose kernel is trivial for the saturation of bosonic
zero modes (this is also the origin of the notion of ``picture'').
However, it uses the non-minimal formulation
of pure-spinor string theory~\cite{Berkovits:2005bt}. Presumably, it might
be related to the present formalism if that would be extended along the
lines of~\cite{Berkovits:2009gi,Kroyter:2009zj}. It would be very
interesting to pursue this direction of research, as it might lead to a
unified picture of superstring field theory, while clarifying some
fundamental issues along the way. A potential obstacle is the fact that the
pure-spinor string field should be allowed to be singular with respect to the
pure-spinor $\la^\al$,
but not too singular. This state of affairs seems to lead to a contradiction
that could only be resolved by modifying the pure-spinor formalism
itself~\cite{Aisaka:2008vw,Bedoya:2009np}. A modification of the pure-spinor
formalism should also allow the introduction of a GSO$(-)$ sector.
One approach towards the inclusion of this sector was presented
in~\cite{Berkovits:2007wz}, where some non-minimal sectors were added to the
theory. However, it is not clear how to unify these non-minimal sectors
with the non-minimal sector of~\cite{Berkovits:2005bt}. Presumably, the
comparison with the democratic theory might give us some clues regarding the
resolution of these difficulties.

We hope that the democratic theory would be found useful for the
construction of closed RNS string field theories. Complications
in generalizing this work are bound to arise, among other reasons, since
in closed string theory the relative cohomologies at different picture
numbers are not all isomorphic~\cite{Berkovits:1997mc}. Presumably, one could
devise a way around this problem by replacing the $b_0^-\Psi=0$
condition by some sort of a gauge symmetry, in a way analogous to the
treatment of $b_0$ in the open string case and the treatment of picture
number in this work. Another potential difficulty comes from the fact that
in the closed string case the mid-point, on which we inserted the operator
$\cO$, is absent. This difficulty could possibly be resolved along the
lines of~\cite{Saroja:1992vw}. This construction could potentially be
generalized, using some ideas of the democratic theory and of the
NS heterotic string field theory~\cite{Berkovits:2004xh}, in order to
construct the desired closed RNS string field theories as well as a
complete RNS heterotic string field theory.
This important issue remains for future work.

Identifying correctly the space of string fields is still an open problem
even within the context of bosonic string field theory. In fact, some of the
most important recent achievements of the field, such as proving Sen's
conjectures~\cite{Sen:1999mh,Sen:1999xm} for Schnabl's
solution~\cite{Schnabl:2005gv} depend crucially on properties of this
unknown space~\cite{Okawa:2006vm,Fuchs:2006hw,Ellwood:2006ba}.
Thus, our proposal that some properties hold for this new theory,
provided the space of string fields has some
given properties, are along the line of what is currently the common
practice. Finding sound definitions for the spaces of string fields of the
various string field theories would be highly important, both for
understanding these theories, as well as for their proper definition.

\section*{Acknowledgments}

I would like to thank Ido Adam, Ofer Aharony, Nathan Berkovits, Stefan
Fredenhagen, Udi Fuchs, Michael Kiermaier, Carlo Maccaferri, Yaron Oz,
Stefan Theisen, Scott Yost and Barton Zwiebach for discussions.

This work is supported by the U.S. Department of Energy
(D.O.E.) under cooperative research agreement DE-FG0205ER41360.
My research is supported by an Outgoing International Marie Curie
Fellowship of the European Community. The views presented in this work are
those of the author and do not necessarily reflect those of the European
Community.

\newpage
\bibliography{bib}

\providecommand{\href}[2]{#2}\begingroup\raggedright\begin{thebibliography}{10}

\bibitem{Friedan:1985ge}
D.~Friedan, E.~J. Martinec, and S.~H. Shenker, {\it Conformal invariance,
  supersymmetry and string theory},  {\em Nucl. Phys.} {\bf B271} (1986) 93.

\bibitem{Witten:1986cc}
E.~Witten, {\it Noncommutative geometry and string field theory},  {\em Nucl.
  Phys.} {\bf B268} (1986) 253.

\bibitem{Witten:1986qs}
E.~Witten, {\it Interacting field theory of open superstrings},  {\em Nucl.
  Phys.} {\bf B276} (1986) 291.

\bibitem{Wendt:1987zh}
C.~Wendt, {\it Scattering amplitudes and contact interactions in {W}itten's
  superstring field theory},  {\em Nucl. Phys.} {\bf B314} (1989) 209.

\bibitem{Preitschopf:1989fc}
C.~R. Preitschopf, C.~B. Thorn, and S.~A. Yost, {\it Superstring field theory},
   {\em Nucl. Phys.} {\bf B337} (1990) 363--433.

\bibitem{Arefeva:1989cm}
I.~Y. Arefeva, P.~B. Medvedev, and A.~P. Zubarev, {\it Background formalism for
  superstring field theory},  {\em Phys. Lett.} {\bf B240} (1990) 356--362.

\bibitem{Arefeva:1989cp}
I.~Y. Arefeva, P.~B. Medvedev, and A.~P. Zubarev, {\it New representation for
  string field solves the consistency problem for open superstring field
  theory},  {\em Nucl. Phys.} {\bf B341} (1990) 464--498.

\bibitem{Berkovits:1995ab}
N.~Berkovits, {\it Super-{P}oincar\'e invariant superstring field theory},
  {\em Nucl. Phys.} {\bf B450} (1995) 90--102,
  [\href{http://xxx.lanl.gov/abs/hep-th/9503099}{{\tt hep-th/9503099}}].

\bibitem{Berkovits:2000hf}
N.~Berkovits, A.~Sen, and B.~Zwiebach, {\it Tachyon condensation in superstring
  field theory},  {\em Nucl. Phys.} {\bf B587} (2000) 147--178,
  [\href{http://xxx.lanl.gov/abs/hep-th/0002211}{{\tt hep-th/0002211}}].

\bibitem{Berkovits:2001im}
N.~Berkovits, {\it The {R}amond sector of open superstring field theory},  {\em
  JHEP} {\bf 11} (2001) 047,
  [\href{http://xxx.lanl.gov/abs/hep-th/0109100}{{\tt hep-th/0109100}}].

\bibitem{Arefeva:2002mb}
I.~Y. Arefeva, D.~M. Belov, and A.~A. Giryavets, {\it Construction of the
  vacuum string field theory on a non-{BPS} brane},  {\em JHEP} {\bf 09} (2002)
  050, [\href{http://xxx.lanl.gov/abs/hep-th/0201197}{{\tt hep-th/0201197}}].

\bibitem{Michishita:2004by}
Y.~Michishita, {\it A covariant action with a constraint and {F}eynman rules
  for fermions in open superstring field theory},  {\em JHEP} {\bf 01} (2005)
  012, [\href{http://xxx.lanl.gov/abs/hep-th/0412215}{{\tt hep-th/0412215}}].

\bibitem{Berkovits:2005bt}
N.~Berkovits, {\it Pure spinor formalism as an {N = 2} topological string},
  {\em JHEP} {\bf 10} (2005) 089,
  [\href{http://xxx.lanl.gov/abs/hep-th/0509120}{{\tt hep-th/0509120}}].

\bibitem{Berkovits:2009gi}
N.~Berkovits and W.~Siegel, {\it Regularizing cubic open {Neveu-Schwarz} string
  field theory},  {\em JHEP} {\bf 11} (2009) 021,
  [\href{http://xxx.lanl.gov/abs/0901.3386}{{\tt 0901.3386}}].

\bibitem{Kroyter:2009zj}
M.~Kroyter, {\it On string fields and superstring field theories},  {\em JHEP}
  {\bf 08} (2009) 044, [\href{http://xxx.lanl.gov/abs/0905.1170}{{\tt
  0905.1170}}].

\bibitem{Kroyter:2009zi}
M.~Kroyter, {\it Superstring field theory equivalence: {Ramond} sector},  {\em
  JHEP} {\bf 10} (2009) 044, [\href{http://xxx.lanl.gov/abs/0905.1168}{{\tt
  0905.1168}}].

\bibitem{Kroyter:2009bg}
M.~Kroyter, {\it Comments on superstring field theory and its vacuum solution},
   {\em JHEP} {\bf 08} (2009) 048,
  [\href{http://xxx.lanl.gov/abs/0905.3501}{{\tt 0905.3501}}].

\bibitem{Fuchs:2008cc}
E.~Fuchs and M.~Kroyter, {\it Analytical solutions of open string field
  theory},  \href{http://xxx.lanl.gov/abs/0807.4722}{{\tt 0807.4722}}.

\bibitem{DeSmet:2000je}
P.-J. De~Smet and J.~Raeymaekers, {\it The tachyon potential in {W}itten's
  superstring field theory},  {\em JHEP} {\bf 08} (2000) 020,
  [\href{http://xxx.lanl.gov/abs/hep-th/0004112}{{\tt hep-th/0004112}}].

\bibitem{Berkovits:2001us}
N.~Berkovits, {\it Relating the {RNS} and pure spinor formalisms for the
  superstring},  {\em JHEP} {\bf 08} (2001) 026,
  [\href{http://xxx.lanl.gov/abs/hep-th/0104247}{{\tt hep-th/0104247}}].

\bibitem{SchnablGrassi}
P.~A. Grassi and M.~Schnabl, {\it to appear}, .

\bibitem{Erler:2007xt}
T.~Erler, {\it Tachyon vacuum in cubic superstring field theory},  {\em JHEP}
  {\bf 01} (2008) 013, [\href{http://xxx.lanl.gov/abs/0707.4591}{{\tt
  0707.4591}}].

\bibitem{Aref'eva:2008ad}
I.~Y. Aref'eva, R.~V. Gorbachev, and P.~B. Medvedev, {\it Tachyon solution in
  cubic {Neveu-Schwarz} string field theory},  {\em Theor. Math. Phys.} {\bf
  158} (2009) 320--332, [\href{http://xxx.lanl.gov/abs/0804.2017}{{\tt
  0804.2017}}].

\bibitem{Fuchs:2008zx}
E.~Fuchs and M.~Kroyter, {\it On the classical equivalence of superstring field
  theories},  {\em JHEP} {\bf 10} (2008) 054,
  [\href{http://xxx.lanl.gov/abs/0805.4386}{{\tt 0805.4386}}].

\bibitem{Berkovits:1994vy}
N.~Berkovits and C.~Vafa, {\it {N=4 topological strings}},  {\em Nucl. Phys.}
  {\bf B433} (1995) 123--180,
  [\href{http://xxx.lanl.gov/abs/hep-th/9407190}{{\tt hep-th/9407190}}].

\bibitem{Schnabl:2005gv}
M.~Schnabl, {\it Analytic solution for tachyon condensation in open string
  field theory},  {\em Adv. Theor. Math. Phys.} {\bf 10} (2006) 433--501,
  [\href{http://xxx.lanl.gov/abs/hep-th/0511286}{{\tt hep-th/0511286}}].

\bibitem{Ellwood:2006ba}
I.~Ellwood and M.~Schnabl, {\it Proof of vanishing cohomology at the tachyon
  vacuum},  {\em JHEP} {\bf 02} (2007) 096,
  [\href{http://xxx.lanl.gov/abs/hep-th/0606142}{{\tt hep-th/0606142}}].

\bibitem{Ellwood:2001ig}
I.~Ellwood, B.~Feng, Y.-H. He, and N.~Moeller, {\it The identity string field
  and the tachyon vacuum},  {\em JHEP} {\bf 07} (2001) 016,
  [\href{http://xxx.lanl.gov/abs/hep-th/0105024}{{\tt hep-th/0105024}}].

\bibitem{NarganesQuijano:1988gb}
F.~J. Narganes-Quijano, {\it Picture changing operation and {BRST} cohomology
  in superstring field theory},  {\em Phys. Lett.} {\bf B212} (1988) 292.

\bibitem{Acosta:1999hi}
J.~N. Acosta, N.~Berkovits, and O.~Chandia, {\it A note on the superstring
  {BRST} operator},  {\em Phys. Lett.} {\bf B454} (1999) 247--248,
  [\href{http://xxx.lanl.gov/abs/hep-th/9902178}{{\tt hep-th/9902178}}].

\bibitem{Callan:1988wz}
C.~G. Callan, Jr., C.~Lovelace, C.~R. Nappi, and S.~A. Yost, {\it Loop
  corrections to superstring equations of motion},  {\em Nucl. Phys.} {\bf
  B308} (1988) 221.

\bibitem{Kiermaier:2008qu}
M.~Kiermaier, Y.~Okawa, and B.~Zwiebach, {\it The boundary state from open
  string fields},  \href{http://xxx.lanl.gov/abs/0810.1737}{{\tt 0810.1737}}.

\bibitem{Horowitz:1988ip}
G.~T. Horowitz, R.~C. Myers, and S.~P. Martin, {\it {BRST} cohomology of the
  superstring at arbitrary ghost number},  {\em Phys. Lett.} {\bf B218} (1989)
  309.

\bibitem{Lian:1989cy}
B.~H. Lian and G.~J. Zuckerman, {\it {BRST} cohomology of the supervirasoro
  algebras},  {\em Commun. Math. Phys.} {\bf 125} (1989) 301.

\bibitem{Berkovits:1999in}
N.~Berkovits, {\it Quantization of the superstring with manifest {$U(5)$}
  super-poincar\'e invariance},  {\em Phys. Lett.} {\bf B457} (1999) 94--100,
  [\href{http://xxx.lanl.gov/abs/hep-th/9902099}{{\tt hep-th/9902099}}].

\bibitem{ZinnJustin:1974mc}
J.~Zinn-Justin, {\it Renormalization of gauge theories}, . Lectures given at
  Int. Summer Inst. for Theoretical Physics, Jul 29 - Aug 9, 1974, Bonn, West
  Germany.

\bibitem{Batalin:1981jr}
I.~A. Batalin and G.~A. Vilkovisky, {\it Gauge algebra and quantization},  {\em
  Phys. Lett.} {\bf B102} (1981) 27--31.

\bibitem{Batalin:1984jr}
I.~A. Batalin and G.~A. Vilkovisky, {\it Quantization of gauge theories with
  linearly dependent generators},  {\em Phys. Rev.} {\bf D28} (1983)
  2567--2582.

\bibitem{Batalin:1984ss}
I.~A. Batalin and G.~A. Vilkovisky, {\it Closure of the gauge algebra,
  generalized {L}ie equations and {F}eynman rules},  {\em Nucl. Phys.} {\bf
  B234} (1984) 106--124.

\bibitem{Voronov:1982cp}
B.~L. Voronov and I.~V. Tyutin, {\it Formulation of gauge theories of general
  form. {I}},  {\em Theor. Math. Phys.} {\bf 50} (1982) 218--225.

\bibitem{Henneaux:1989jq}
M.~Henneaux, {\it Lectures on the {Antifield-BRST} formalism for gauge
  theories},  {\em Nucl. Phys. Proc. Suppl.} {\bf 18A} (1990) 47--106.

\bibitem{Henneaux:1992ig}
M.~Henneaux and C.~Teitelboim, {\it Quantization of gauge systems}, .
  Princeton, USA: Univ. Pr. (1992) 520 p.

\bibitem{Gomis:1994he}
J.~Gomis, J.~Paris, and S.~Samuel, {\it Antibracket, antifields and gauge
  theory quantization},  {\em Phys. Rept.} {\bf 259} (1995) 1--145,
  [\href{http://xxx.lanl.gov/abs/hep-th/9412228}{{\tt hep-th/9412228}}].

\bibitem{Thorn:1986qj}
C.~B. Thorn, {\it Perturbation theory for quantized string fields},  {\em Nucl.
  Phys.} {\bf B287} (1987) 61.

\bibitem{Bochicchio:1986bd}
M.~Bochicchio, {\it String field theory in the {S}iegel gauge},  {\em Phys.
  Lett.} {\bf B188} (1987) 330.

\bibitem{Bochicchio:1986zj}
M.~Bochicchio, {\it Gauge fixing for the field theory of the bosonic string},
  {\em Phys. Lett.} {\bf B193} (1987) 31.

\bibitem{Thorn:1988hm}
C.~B. Thorn, {\it String field theory},  {\em Phys. Rept.} {\bf 175} (1989)
  1--101.

\bibitem{D'Hoker:2002gw}
E.~D'Hoker and D.~H. Phong, {\it Lectures on two-loop superstrings},
  \href{http://xxx.lanl.gov/abs/hep-th/0211111}{{\tt hep-th/0211111}}.

\bibitem{Morozov:2008wz}
A.~Morozov, {\it {NSR} superstring measures revisited},  {\em JHEP} {\bf 05}
  (2008) 086, [\href{http://xxx.lanl.gov/abs/0804.3167}{{\tt 0804.3167}}].

\bibitem{DuninBarkowski:2009ej}
P.~Dunin-Barkowski, A.~Morozov, and A.~Sleptsov, {\it Lattice theta constants
  vs {Riemann} theta constants and {NSR} superstring measures},  {\em JHEP}
  {\bf 10} (2009) 072, [\href{http://xxx.lanl.gov/abs/0908.2113}{{\tt
  0908.2113}}].

\bibitem{Kiermaier:2007jg}
M.~Kiermaier, A.~Sen, and B.~Zwiebach, {\it Linear b-gauges for open string
  fields},  {\em JHEP} {\bf 03} (2008) 050--050,
  [\href{http://xxx.lanl.gov/abs/0712.0627}{{\tt 0712.0627}}].

\bibitem{Asano:2006hk}
M.~Asano and M.~Kato, {\it New covariant gauges in string field theory},  {\em
  Prog. Theor. Phys.} {\bf 117} (2007) 569--587,
  [\href{http://xxx.lanl.gov/abs/hep-th/0611189}{{\tt hep-th/0611189}}].

\bibitem{Asano:2008iu}
M.~Asano and M.~Kato, {\it General linear gauges and amplitudes in open string
  field theory},  {\em Nucl. Phys.} {\bf B807} (2009) 348--372,
  [\href{http://xxx.lanl.gov/abs/0807.5010}{{\tt 0807.5010}}].

\bibitem{Kroyter:2010rk}
M.~Kroyter, {\it Democratic superstring field theory: Gauge fixing},
  \href{http://xxx.lanl.gov/abs/1010.1662}{{\tt 1010.1662}}.

\bibitem{Berkovits:1996bf}
N.~Berkovits, {\it A new description of the superstring},
  \href{http://xxx.lanl.gov/abs/hep-th/9604123}{{\tt hep-th/9604123}}.

\bibitem{Okawa:2006vm}
Y.~Okawa, {\it Comments on {S}chnabl's analytic solution for tachyon
  condensation in {W}itten's open string field theory},  {\em JHEP} {\bf 04}
  (2006) 055, [\href{http://xxx.lanl.gov/abs/hep-th/0603159}{{\tt
  hep-th/0603159}}].

\bibitem{Fuchs:2007yy}
E.~Fuchs, M.~Kroyter, and R.~Potting, {\it Marginal deformations in string
  field theory},  {\em JHEP} {\bf 09} (2007) 101,
  [\href{http://xxx.lanl.gov/abs/arXiv:0704.2222 [hep-th]}{{\tt arXiv:0704.2222
  [hep-th]}}].

\bibitem{Fuchs:2007gw}
E.~Fuchs and M.~Kroyter, {\it Marginal deformation for the photon in
  superstring field theory},  {\em JHEP} {\bf 11} (2007) 005,
  [\href{http://xxx.lanl.gov/abs/arXiv:0706.0717 [hep-th]}{{\tt arXiv:0706.0717
  [hep-th]}}].

\bibitem{Qiu:1987dv}
Z.-a. Qiu and A.~Strominger, {\it Gauge symmetries in (super)string field
  theory},  {\em Phys. Rev.} {\bf D36} (1987) 1794.

\bibitem{Berkovits:2000fe}
N.~Berkovits, {\it {Super-Poincare} covariant quantization of the superstring},
   {\em JHEP} {\bf 04} (2000) 018,
  [\href{http://xxx.lanl.gov/abs/hep-th/0001035}{{\tt hep-th/0001035}}].

\bibitem{Berkovits:2002zk}
N.~Berkovits, {\it {ICTP} lectures on covariant quantization of the
  superstring},  \href{http://xxx.lanl.gov/abs/hep-th/0209059}{{\tt
  hep-th/0209059}}.

\bibitem{Aisaka:2008vw}
Y.~Aisaka, E.~A. Arroyo, N.~Berkovits, and N.~Nekrasov, {\it Pure spinor
  partition function and the massive superstring spectrum},  {\em JHEP} {\bf
  08} (2008) 050, [\href{http://xxx.lanl.gov/abs/0806.0584}{{\tt 0806.0584}}].

\bibitem{Bedoya:2009np}
O.~A. Bedoya and N.~Berkovits, {\it {GGI} lectures on the pure spinor formalism
  of the superstring},  \href{http://xxx.lanl.gov/abs/0910.2254}{{\tt
  0910.2254}}.

\bibitem{Berkovits:2007wz}
N.~Berkovits, {\it Explaining the pure spinor formalism for the superstring},
  {\em JHEP} {\bf 01} (2008) 065,
  [\href{http://xxx.lanl.gov/abs/0712.0324}{{\tt 0712.0324}}].

\bibitem{Berkovits:1997mc}
N.~Berkovits and B.~Zwiebach, {\it On the picture dependence of {Ramond-Ramond}
  cohomology},  {\em Nucl. Phys.} {\bf B523} (1998) 311--343,
  [\href{http://xxx.lanl.gov/abs/hep-th/9711087}{{\tt hep-th/9711087}}].

\bibitem{Saroja:1992vw}
R.~Saroja and A.~Sen, {\it Picture changing operators in closed fermionic
  string field theory},  {\em Phys. Lett.} {\bf B286} (1992) 256--264,
  [\href{http://xxx.lanl.gov/abs/hep-th/9202087}{{\tt hep-th/9202087}}].

\bibitem{Berkovits:2004xh}
N.~Berkovits, Y.~Okawa, and B.~Zwiebach, {\it {WZW}-like action for heterotic
  string field theory},  {\em JHEP} {\bf 11} (2004) 038,
  [\href{http://xxx.lanl.gov/abs/hep-th/0409018}{{\tt hep-th/0409018}}].

\bibitem{Sen:1999mh}
A.~Sen, {\it Descent relations among bosonic {D}-branes},  {\em Int. J. Mod.
  Phys.} {\bf A14} (1999) 4061--4078,
  [\href{http://xxx.lanl.gov/abs/hep-th/9902105}{{\tt hep-th/9902105}}].

\bibitem{Sen:1999xm}
A.~Sen, {\it Universality of the tachyon potential},  {\em JHEP} {\bf 12}
  (1999) 027, [\href{http://xxx.lanl.gov/abs/hep-th/9911116}{{\tt
  hep-th/9911116}}].

\bibitem{Fuchs:2006hw}
E.~Fuchs and M.~Kroyter, {\it On the validity of the solution of string field
  theory},  {\em JHEP} {\bf 05} (2006) 006,
  [\href{http://xxx.lanl.gov/abs/hep-th/0603195}{{\tt hep-th/0603195}}].

\end{thebibliography}\endgroup

\end{document}